\RequirePackage{fix-cm}
\documentclass[twocolumn]{svjour3}          
\usepackage{orcidlink}
\usepackage{multirow}
\usepackage[misc]{ifsym}
\usepackage{tcolorbox}
\usepackage{amsmath}
\usepackage{enumitem}
\usepackage{mdframed}
\usepackage{xcolor}
\usepackage{subcaption}
\usepackage{url}
\usepackage[ruled, vlined, linesnumbered]{algorithm2e}
\usepackage{setspace}
\SetAlgoNlRelativeSize{-2} 
\usepackage{appendix}
\usepackage{changepage}
\usepackage{calc}
\usepackage{makecell}
\smartqed  

\usepackage{xspace}
\usepackage{ifthen}
\usepackage{marginnote}

\newcommand{\bao}[1]{}
\newcommand{\daomin}[1]{}
\newcommand{\shane}[1]{}
\newcommand{\myparagraph}[1]{\vspace{0.3\baselineskip}\noindent{\textbf{#1.}}~}
\newcommand{\response}[1]{\vspace{0.3\baselineskip}\noindent{Response.~}}

\newcommand{\add}[1]{{\color{black}{#1}}}

\newcommand{\methodone}{SSACL\xspace}
\newcommand{\methodtwo}{ICLCR\xspace}
\newcommand{\slim}{SlimFast\xspace}
\newcommand{\rand}{Random\xspace}
\newcommand{\major}{Major\xspace}

\newcommand{\taska}{pairwise integrability judgment\xspace}
\newcommand{\taskb}{integrable set discovery\xspace}
\newcommand{\taskc}{multi-tuple conflict resolution\xspace}

\newcommand{\Taska}{Pairwise integrability judgment\xspace}
\newcommand{\Taskb}{Integrable set discovery\xspace}
\newcommand{\Taskc}{Multi-tuple conflict resolution\xspace}

\newcommand{\TASKA}{Pairwise Integrability Judgment\xspace}
\newcommand{\TASKB}{Integrable Set Discovery\xspace}
\newcommand{\TASKC}{Multi-tuple Conflict Resolution\xspace}

\newcommand{\generator}{data generator\xspace}
\newcommand{\encoder}{encoder\xspace}
\newcommand{\matcher}{matcher\xspace}
\newcommand{\trainer}{adversial trainer\xspace}

\begin{document}

\title{Table Integration in Data Lakes Unleashed:  Pairwise Integrability Judgment, Integrable Set Discovery, and Multi-Tuple Conflict Resolution
}


\titlerunning{Table Integration in Data Lakes Unleashed}        

\author{Daomin Ji\textsuperscript{1}    \and
        Hui Luo\textsuperscript{2}  \and
        Zhifeng Bao\textsuperscript{1} \and
        J. Shane Culpepper\textsuperscript{3}
}

\institute{
Zhifeng Bao~\Letter \\
zhifeng.bao@rmit.edu.au \\ \\
Daomin Ji \\
daomin.Ji@student.rmit.edu.au \\ \\
Hui Luo \\
huil@uow.edu.au \\ \\ 
J. Shane Culpepper\\  
s.culpepper@uq.edu.au\\ \\ 
\textsuperscript{1} RMIT University, Melbourne, Australia \\ \\ 
\textsuperscript{2} University of Wollongong, Wollongong, Australia  \\ \\
\textsuperscript{3} The University of Queensland, Brisbane, Australia
}

\date{Received: date / Accepted: date}

\maketitle

\begin{abstract}
Table integration aims to create a comprehensive table by consolidating tuples containing relevant information.
In this work, we investigate the challenge of integrating multiple tables from a data lake, focusing on three core tasks: 1) \emph{\taska}, which determines whether a tuple pair is integrable, accounting for any occurrences of semantic equivalence or typographical errors; 
2) \emph{\taskb}, which identifies all integrable sets in a table based on pairwise integrability judgments established in the first task; 
3) \emph{\taskc}, which resolves conflicts between multiple tuples during integration. 
To this end, we train a binary classifier to address the task of \taska.
Given the scarcity of labeled data in data lakes, we propose a self-supervised adversarial contrastive learning algorithm to perform classification, which incorporates data augmentation methods and adversarial examples to autonomously generate new training data. 
Upon the output of \taska, each integrable set can be considered as a community—a densely connected sub-graph where nodes and edges correspond to tuples in the table and their pairwise integrability, respectively—we proceed to investigate various community detection algorithms to address the \taskb objective.
Moving forward to tackle \emph{\taskc}, we introduce an innovative in-context learning methodology. This approach capitalizes on the knowledge embedded within large language models (LLMs) to effectively resolve conflicts that arise when integrating multiple tuples. Notably, our method minimizes the need for annotated data, making it particularly suited for scenarios where labeled datasets are scarce. 
Since no suitable test collections are available for our tasks, we develop our own benchmarks using two real-word dataset repositories:  \emph{Real} and \emph{Join}. We conduct extensive experiments on these benchmarks to validate the robustness and applicability of our methodologies in the context of integrating tables within data lakes. 
\end{abstract}

\section{Introduction} \label{sec1: intro}
Data lakes are large repositories that store various types of raw data~\cite{vldb_data_lake_management_nargesian2019data,constance_hai2016constance}. 
Recently, there has been a growing interest in performing table discovery tasks~\cite{santos_khatiwada2023santos,starmie_fan2022semantics,fan2023tabletsnew,zhu2019josietsnew,nargesian2018tabletsnew,cappuzzo2024retrieve} to find unionable, joinable or similar tables in large data lakes. 
The integration of data lake tables into a more unified and comprehensive table can potentially be used to create new knowledge and insights that would otherwise be inaccessible from using the tables in isolation.
Specifically, given a set of input tables from data lakes, the objective is to produce a comprehensive table by merging relevant tuples from different tables into unified tuples. 
To enable table integration in data lakes, four core tasks must be resolved: 
\begin{itemize}[leftmargin=*,noitemsep] 
\item \textit{Schema alignment.} Given two sets of attributes from the tables, the goal is to learn a mapping that aligns each attribute in one set to its corresponding attribute in the other. 
\item \textit{\Taska.} For any two tuples, whether from the same table or different ones, determine if they should be integrated.
\item \textit{\Taskb.}  Based on the judged pairwise integrability, identify all integrable sets across the tables, which indicate exactly which tuples should be integrated together. 
\item \textit{\Taskc.} For each integrable set, produce a single tuple that consolidates all relevant information by reconciling any attribute-level conflicts. 
\end{itemize}

Note that \taska is similar to entity resolution~\cite{dadder_tu2022domain,deeper_ebraheem2018distributed,khatiwada2025fuzzy}, which aims to determine whether two tuples refer to the same entity.
However, in our context, two tuples might be integrated even if they do not strictly correspond to the same entity.
For example, a tuple representing a movie and another representing a director may be merged into a new tuple for the director, where the movie becomes an attribute of the director tuple. 
We also note that existing studies have explored integrating tuples using schema-agnostic entity resolution methods~\cite{simonini2018schemaagnostic1,teong2020schemaagnostic2}, which bypass the schema alignment step. However, in our case, schema-aware methods are preferred and the schema alignment is a prerequisite step, as our objective is to produce a comprehensive table with a unified schema. The aligned schema not only supports this goal but also facilitates subsequent tasks.

\begin{figure*}[pht]
    \centering
    \includegraphics[width=\linewidth]{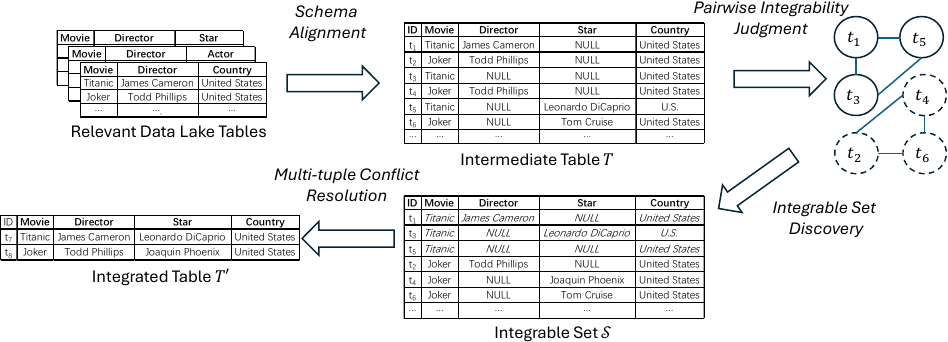}
    \caption{A workflow example for data lake table integration}\label{fig: workflow}
\end{figure*}

\begin{example}
Fig.~\ref{fig: workflow} demonstrates an example of table integration. First, the schemas of all tables must be aligned. While most attributes, such as Movie, Country, and Director, can be directly aligned, the schema alignment method needs to recognize that Actor and Star refer to the same attribute. Next, the tables are combined into an intermediate table $T$ using an outer union operator. During the pairwise integrability judgment phase, it is necessary to evaluate whether any tuple pair can be integrated, even when the tuples contain semantically equivalent values or typographical errors. For example, in tuples $t_1$ and $t_5$, ``U.S.'' and ``United States'' represent semantically equivalent values, whereas in tuples $t_4$ and $t_6$, ``United Skates'' is a typographical error of ``United States.'' These cases must be carefully addressed during the pairwise integrability judgment. Based on the pairwise integrability of the tuples, the table $T$ is partitioned into two integrable sets: one related to the movie Titanic and the other to Joker. Finally, the tuples within each integrable set are integrated into a single comprehensive tuple through multi-tuple conflict resolution. Since conflicting values may exist within an attribute in the same integrable set (e.g., ``Joaquin Phoenix''and ``Tom Cruise'' in the Star attribute), the correct value, such as ``Joaquin Phoenix,'' is selected to produce the final tuple.
\end{example}

Given the high accuracy of existing schema alignment methods, this work focuses on the latter three tasks. Thus, we assume the schemas of all input tables have been aligned by existing schema alignment methods, and the input tables are combined into an intermediate table $T$. Specifically, integrating tables from data lakes presents the following challenges to these tasks:

\noindent \textit{Pairwise integrability judgment.} 
Compared to relational tables, data lake tables~\cite{vldb_data_lake_management_nargesian2019data} often contain significant amounts of dirty data, such as typographical errors and missing values (present in the original input tables or introduced through the outer union operator), in addition to semantically equivalent values. As a result, pairwise integrability judgment for data lake tables requires more robust methods to handle these challenges effectively. Furthermore, most existing pairwise integrability judgment methods rely on large amounts of labeled data to train machine learning models~\cite{deeper_ebraheem2018distributed,deepmatcher_mudgal2018deep}. However, in the context of data lakes, labeled data is often scarce, making it difficult to apply these methods without addressing the label scarcity issue.

\noindent \textit{Integrable set discovery.} Data lake tables may contain massive amounts of dirty tuples that need to be integrated, necessitating a robust and effective approach to accurately identify all integrable sets.

\noindent \textit{Multi-tuple conflict resolution.} Existing conflict resolution solutions often rely on truth discovery methods, which estimate the reliability of each data source and select the source with the highest reliability score. However, these methods typically require labeled data, metadata (e.g., citations or page views), or domain knowledge. Such resources are scarce in data lakes, motivating the development of a conflict resolution method effective with limited labeled data.

In addressing the task of \textit{\taska}, our objective is to accurately predict the integrability of each tuple pair within a table $T$, even in the presence of semantic equivalence and typographical errors. Our solution starts from training a binary classifier, where a major challenge lies in the scarcity of labeled data specific to data lake tables.
To mitigate this challenge, we adopt a strategy where semantic equivalence and typographical errors are treated as minor perturbations of each tuple $t$.  We then employ data augmentation techniques along with adversarial examples to simulate these perturbations. 
This allows us to automatically generate a sufficient amount of training data to train a binary classifier, thereby overcoming the limitation imposed by the scarcity of labeled data.

For the task of \textit{\taskb}, 
we address it from two perspectives: 1) Each integrable set can be considered as a maximal clique in a graph constructed from table $T$, and hence, our objective is to find maximal cliques in an undirected graph.
Thus, we propose to employ the well-known Bron-Kerbosch algorithm~\cite{regneri2007bronalg} to find these integrable sets. 
2) Recognizing that predictions from \taska may contain errors, leading integrable sets may not strictly conform to a clique structure. Hence, we relax the connectivity criteria and view an integrable set as a densely connected subgraph, akin to a community. To handle this, we propose to employ community detection methods tailored for identifying such structures.

Finally, to solve the task of \textit{\taskc}, we
depart from conventional conflict resolution approaches
in data fusion~\cite{vldb_df_dong2009data,vldb_df_li2014confidence}, which typically rely on extensive labeled data.
Instead, we propose a novel method called in-context learning for conflict resolution (\methodtwo).
This method leverages the extensive knowledge embedded in large language models (LLMs), which require only a few labeled demonstration examples to predict conflict resolution outcomes. 
This approach significantly reduces the dependency on large labeled datasets while maintaining comparable performance level. 
The effectiveness of our approach is closely linked to the number and quality of the demonstration examples. 
However, the input size limitations of an LLM limit the number of demonstration examples that can be used. 
To overcome this limitation, we propose an effective strategy for compressing demonstration examples, reducing the average number of tokens to represent an example, thereby enabling inclusion of more examples in a single input.  
Additionally, we introduce targeted strategies for selecting demonstration examples that are most relevant to the conflict resolution task, further enhancing the overall performance of our model.

In summary, our approach exhibits significant potential to enhance table integration within data lakes, offering the following contributions: 

\begin{itemize}[noitemsep, leftmargin=*]
\item To solve the task of \taska, we propose a novel \underline{S}elf-\underline{S}upervised \underline{A}dversarial \underline{C}ontr-
astive \underline{L}earning framework, \methodone, to train a binary classifier with limited labeled data, to predict pairwise integrability of tuple pairs (Sec.~\ref{sec: integration condition judgment}).
\item To solve the task of \taskb, we propose two different yet related approaches. We explore existing solutions relevant to these problems to effectively support the task of \taskb (Sec.~\ref{sec: integrable set identification}). 
\item To solve the task of \taskc, we develop an in-context learning-based method, namely \underline{I}n-\underline{c}ontext \underline{L}earning for \underline{C}onflict \underline{R}esolution (\methodtwo), which demonstrates promising performance with limited labeled data (Sec.~\ref{sec: multi-tuple integration}).
\item Since no suitable benchmarks exist to evaluate our problem, we have taken the initiative to create our owns and make it public~\cite{github} (Sec.~\ref{sec: data preparation}).
\item We conduct an extensive evaluation of our methods and compare it against suitable baselines on the new benchmarks.
When comparing against the best-performing competitors, our \methodone exhibits a relative improvement of $4.2\%$ in terms of \emph{F1} on the task of \taska, and \methodtwo achieves a relative improvement of $18.9\%$ in \emph{Accuracy} on the task of \taskc. 
Furthermore, both \methodone and \methodtwo, when using limited labeled data, experience a decrease in performance of less than $10\%$, compared with when they are trained with a sufficient amount of labeled data.
We also find that among all the methods proposed for \taskb, Graph Neural Network (GNN) achieves the best performance (Sec.~\ref{sec: experiment}).
\end{itemize}

\section{Problem Formulation}\label{sec: problem formulation}
Given a set of tables retrieved from data lakes, table integration seeks to consolidate these tables into a comprehensive unified table by merging relevant tuples from different sources. As discussed in Sec.~\ref{sec1: intro}, this process involves four core tasks: \textit{schema alignment}, \textit{\taska}, \textit{\taskb}, and \textit{\taskc}. Since the schema alignment task has been extensively studied and existing methods can produce highly accurate results, this work focuses on the latter three tasks. Thus, we assume that all tables undergoing integration have already been schema-aligned and combined into an intermediate table $T$ using an outer union operator.
We now introduce the problem definition of \taska.

\begin{definition}[\textbf{\Taska.}]~\label{def: integration condition judgment}
Given a tuple pair $(t_i, t_j)$, which may be semantically equivalent or contain typographical errors, the goal of pairwise integrability judgment, denoted as a function $f$, is to determine whether the tuples are integrable. Specifically, $f(t_i, t_j) = 1$ if $t_i$ and $t_j$ are integrable; otherwise, $f(t_i, t_j) = 0$.
\end{definition}

After obtaining the pairwise integrability for all tuple pairs in table $T$, the next task is to identify all of the integrable sets in $T$, as defined in Def.~\ref{def: integrable set discovery}.

\begin{definition}[\textbf{\Taskb.}]\label{def: integrable set discovery}
Given a table $T$ and a pairwise integrability judgment function $f$, multiple disjoint integrable sets may be produced, denoted as $\mathcal{S}=\{S_1, S_2, ..., S_{|\mathcal{S}|}\}$. 
Each integrable set $S$ ($S \in \mathcal{S}$) contains a collection of tuples that meet two conditions:
\begin{itemize}[leftmargin=*]
\item \textbf{Consistency.} For any two tuples $t_i \in S, t_j \in S$, $f(t_i,t_j)=1$ should always hold.
\item \textbf{Maximality.} For each tuple $t_i \notin S$, there should exist at least one tuple $t_j \in S$ such that $f(t_i, t_j)=0$.
\end{itemize}
\end{definition}

As described in Definition~\ref{def: integrable set discovery}, any pair of tuples in the integrable set should be integrable.
However, given that the output from the method of \taska is not perfect, we relax the definition of an integrable set as follows:

\begin{definition}[\textbf{\Taskb (Relaxed Version)}]\label{def: relaxed integrable set discovery}
Given a table $T$ and a pairwise integrability judgment function $f$, multiple disjoint integrable sets may be produced, denoted as $\mathcal{S}={S_1, S_2, ..., S_{|\mathcal{S}|}}$.
Each integrable set $S$ ($S \in \mathcal{S}$) contains a collection of tuples such that each tuple is integrable with the majority class from the tuples in $S$.  \end{definition}

We will provide a detailed explanation of how the relaxed definition of the integrable set is derived in Sec.~\ref{sec: integrable set identification}.

For each integrable set $S \in \mathcal{S}$, all tuples in $S$ are considered to be integrable into a single tuple, denoted as $t_{new}$.
This integration process involves filling the attributes of $t_{new}$ with the correct value. 
However, when an attribute $t_{new}$ has multiple distinct values that originate from different tuples in $S$, we refer to it as a {\em conflict}, which can be formally addressed by \textit{\taskc}.

\begin{definition}[\textbf{\Taskc.}]\label{def: conflict resolution}
Given an integrable set $S$, a tuple $t_{new}$ is produced by integrating all tuples in $S$, such that, for each attribute $a$ in $t_{new}$, where a conflict exists, the correct value $v^*$ is chosen from the candidate set $C=\{t[a]|t \in S \land t[a] \neq NULL \}$ to complete the attribute value $t_{new}[a]$.
\end{definition}

\section{\TASKA}\label{sec: integration condition judgment}

\subsection{Key Idea}\label{sec: main idea for task 1}
Given that the pairwise integrability judgment function $f$ can be regarded as a binary classifier, our goal is to use a machine learning model to approximate $f$.
However, a significant hurdle exists due to the unavailability of labeled data in data lake environments, and manually labeling data is prohibitively expensive.

To overcome this hurdle, we propose a novel approach -- \emph{self-supervised adversarial contrastive learning} (\methodone), which is designed to automatically generate training data for both positive and negative instances.Contrastive learning is a machine learning approach that involves distinguishing between similar and dissimilar data points by bringing similar points closer together in the feature space while pushing dissimilar points further apart. This approach has been widely adopted in table representation learning tasks~\cite{starmie_fan2022semantics,brinkmann2024CL1}.
Specifically, for a given tuple $t$, although the negative instances (tuples that cannot be integrated with $t$) can be obtained using negative sampling~\cite{awasthi2022morenegativesampling,armandpour2019robustnegativesampling}, the key challenge is \emph{how to generate positive instances (tuples that can be integrated with $t$)}, particularly tuples that are semantically equivalent to tuple $t$ or exhibit typographical errors.
To address this challenge, given that tuples with typographical errors and semantic equivalence should not differ significantly with the original tuple in the semantics, for each tuple $t$, we introduce a slight perturbation function $p$ to generate a positive instance $t^+$, i.e., $t^+=p(t)$.
This slight perturbation function $p$ is designed to induce minimal semantic divergence from$t$ to $t^+$, effectively simulating semantic equivalence and typographical errors.
Specifically, we employ two strategies to simulate $p$: \emph{data augmentation}~\cite{csur2022da,feng2021dafornlp,wang2023sudowoodo} and \emph{adversarial examples}~\cite{ilyasnlps2019adversarial,zhangtnnls2019adversarial,goodfellowiclr2014explaining}, which will be described in Sec.~\ref{sec: generator} and Sec.~\ref{sec: trainer}, respectively.

Naturally, for a tuple pair $(t,t^+)$, $f(t,t^+)=1$ should hold.
Once the model $f$ has been sufficiently trained on positive instances $(t, t^+)$, accurately inferring the integrability of two tuples is plausible, even when the tuples are semantically equivalent or contain typographical errors.
Furthermore, the binary classifier is trained using the contrastive learning framework, with the objective of producing embeddings where positive tuple pairs are closer together in the embedding space and negative tuple pairs are farther apart. 

\par Fig.~\ref{fig: methodone} presents the architecture for our proposed \methodone, which has the following key components:

\begin{figure*}
    \centering
    \includegraphics[width=\linewidth]{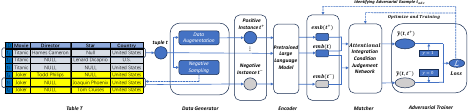}
    \captionsetup{skip=3pt}
    \caption{Architecture of our proposed \methodone}
    \label{fig: methodone}
\end{figure*}

\begin{itemize}[noitemsep, leftmargin=*]
\item \emph{Data generator} employs data augmentation and negative sampling to generate positive instances $t^+$ and negative instances $t^-$ for each tuple $t$, forming a training data set $\mathcal{D}_{train}$ (Sec.~\ref{sec: generator}). 
\item \emph{Encoder} transforms each tuple $t\in \mathcal{D}_{train}$ generated from the \textit{\generator} to a compact and semantically meaningful representation $\boldsymbol{emb}(t)$ (Sec.~\ref{sec: encoder}).
\item \emph{Matcher} is designed to evaluate the compatibility of two tuples. It takes embeddings produced by the \emph{Encoder} as input and outputs either 1 or 0, indicating whether the tuples are integrable or not (Sec.~\ref{sec: matcher}).
\item \emph{Adversarial trainer} is designed to generate additional positive training instances by leveraging adversarial examples. Furthermore, it is employed to optimize the parameters of the \emph{Matcher} (Sec.~\ref{sec: trainer}).
\end{itemize}

\subsection{Data Generator}\label{sec: generator}
A data generator automatically generates training data for \taska. 
Specifically, for each tuple $t$, we use data augmentation techniques to produce a collection of perturbed tuples, $\mathcal{P}_t=\{t^+_i|t^+_i=p_i(t)\}$, where $p_i$ corresponds to a specific perturbation function.
Consequently, every tuple pair $(t,t^+_i)$ is a positive training instance for the \matcher. 
Obviously, deciding how to create the slight perturbations $p_i$ impacts the quality of the training data, which in turn affects the performance of the training model $f$.
So, the selection of perturbation functions should be comprehensive.
Perturbation functions are carefully selected to adhere to two key principles: (1) The perturbations should preserve the semantics of the original tuple $t$ or make minimal changes; (2) The perturbation functions should ideally cover a diverse range of possibilities to effectively represent multiple real-world scenarios.
To achieve this, we develop a variety of perturbation functions which are organized into three types: attribute-level, word-level, and character-level.

\noindent\textbf{Attribute-level.} There are two kinds of attribute-level perturbations:
\begin{itemize}[noitemsep,leftmargin=*]
\item \emph{Attribute removal.} This is applied to both numerical and textual attributes. 
We randomly select a non-null attribute value $a$ from a tuple $t$ and create a new tuple $t^+$ by removing the selected attribute value, such that $t^+[a]==NULL$.
\item \emph{Attribute substitution}. This is applied to textual attributes only. 
We randomly select an attribute $a$ of a tuple $t$, and create a new tuple $t^+$ by using backtranslation  methods~\cite{edunov2018understanding,shorten2021text}, which generate a semantically equivalent description for $t[a]$, and use this new description to replace the original value $t[a]$.
\end{itemize}

\noindent\textbf{Word-level.} There are three types of word-level perturbations adopted in \methodone:

\begin{itemize}[noitemsep,leftmargin=*]
\item \emph{Word Removal.} For a tuple attribute value, we randomly select a word and delete it. 
\item \emph{Word Substitution.} For a tuple attribute value, we randomly select a word, and use WordNet~\cite{miller1995wordnet} to find synonyms or hypernyms to form a candidate set.
Finally, we randomly select a word from the candidate set to substitute the original term.
\item \emph{Word Swapping.} For a tuple attribute value, we randomly select two neighboring words, and swap their positions.
\end{itemize}

\noindent\textbf{Character-level.} For character-level perturbations, we simulate common typographical errors to create new tuples, which can be applied to both textual attributes and numerical attributes.
More details on how to simulate common typographical errors can be found in~\cite{typo_github}.

\par Specifically, for each original tuple $t$, we randomly select $N_{pos}$ perturbation functions to perturb the tuples $t^+$. 
In other words, we create $N_{pos}$ positive training instances for each original tuple $t$, and for negative training instances, we adopt the widely used strategy of negative sampling~\cite{awasthi2022morenegativesampling,armandpour2019robustnegativesampling} to uniformly select $\mathcal{N}_{neg}$ tuples ($t$ to be excluded) at random in the training table for each tuple $t$.

\subsection{Encoder}\label{sec: encoder}
Given a tuple $t$, the encoder learns a compact and semantically meaningful embedding $\boldsymbol{emb}(t)$ to represent $t$. 
In this paper, we adopt attribute-level representations for each tuple, enabling us to perform fine-grained comparisons for the attribute values from any two tuples.
While we are aware of recent work that directly uses pre-trained language models (e.g.,~\cite{unicorn_tu2023unicorn}) to encode serialized rows, this approach produced suboptimal results in our case. A possible reason is that the outer union operation introduces many missing values, which may distract the PLM and affect its performance.
Specifically, for each tuple, $t$, the \emph{encoder} encodes the following representation by combining all attribute-level representations:
\begin{equation}\label{Eq: tuple embedding}
    \boldsymbol{emb}(t)= [\boldsymbol{emb}(t[a_1]), \boldsymbol{emb}(t[a_2]), ..., \boldsymbol{emb}(t[a_m])].
\end{equation}
Here, $m$ denotes the number of attributes in Table $T$, $t[a_i]$ represents the value of the attribute $a_i$ in the tuple $t$, and $\boldsymbol{emb}(t[a_i])$ denotes the corresponding embedding.
Note that for a tuple $t$, there may be missing attributes.
To handle such cases, we use a special token $[NULL]$ to mark missing values, which will also be mapped to an embedding. 
Furthermore, for each tuple $t$, we also create a masking vector $\boldsymbol{Mask}(t)$ that has $m$ dimensions: 
\begin{equation}
    \boldsymbol{Mask}(t) = [d_1, d_2, ..., d_m],
\end{equation}
where we set the $i\cdot$th element of the masking vector $d_i$ to 0 if the corresponding attribute value for tuple $t$ is missing; otherwise, we set it to 1.
The masking vectors are discussed further in Sec.~\ref{sec: matcher}.

\par To obtain $\boldsymbol{emb}(t[a_i])$, we first serialize an attribute value $t[a_i]$ into a sequence of words.
Then, for each word $w$ in the sequence, we generate an embedding using a pre-trained language model.
There are two different embedding methods, namely word-level embeddings such as word2vec~\cite{mikolov2013word2vec} and GloVE~\cite{pennington2014glove}, and subword-level embeddings such as FastText~\cite{bojanowski2017fasttext} and BERT~\cite{kenton2019bert}. 
Word2vec encodes each term individually and uses an embedding vector to represent it, whereas GloVE tokenizes a word into a sequence of subwords and represents each subword using an embedding.
In this work, we use a subword-level embedding method.
This choice is driven by their capability to handle unknown and uncommon terms, while also exhibiting greater resilience to typographical errors.

\par For every sequence of tokens, the respective token embeddings are aggregated into a single embedding vector. We adopt a transformer-based architecture~\cite{kenton2019bert} for this  aggregation process.
This choice stems from their proven ability to adeptly capture contextual information embedded in sequences in the transformer.
As a result, we obtain an embedding $\boldsymbol{emb}(t[a_i])$ tuned for the \matcher, which we will discuss in Section \ref{sec: matcher}. 

\subsection{Matcher}\label{sec: matcher}
Given the embedding representation for two tuples $\boldsymbol{emb}(t_1)$ and $\boldsymbol{emb}(t_2)$, the \matcher outputs 1 or 0, indicating whether the two tuples should be integrated.
One straightforward way to achieve this is to compute the cosine similarity between $\boldsymbol{emb}(t_1)$ and $\boldsymbol{emb}(t_2)$, or concatenate the two embeddings into a Multi-layer Perceptron (MLP).
A common drawback emerges from the uniform treatment of every attribute, which ignores any potential variations in the contributions to the semantic correctness for a given tuple.

\par To overcome this drawback, we propose an \emph{Attentional Integrability Judgment Network} (AIJNet), which assigns different weights to the attributes in the matching process.
Specifically, for two embeddings $\boldsymbol{emb}(t_1)$ and $\boldsymbol{emb}(t_2)$, we first concatenate both into a single embedding: $\boldsymbol{emb}(t_1, t_2) = [\boldsymbol{emb}(t_1),\boldsymbol{emb(t_2)}]$ . 
Next, we consider the varying importance of each attribute and reformulate the representation of the tuple pair $(t_1, t_2)$ using a self-attention mechanism~\cite{vaswani2017attention}: 
\begin{equation}
\boldsymbol{emb}^*(t_1, t_2)_i = \sum_{j=1}^{2m} \text{softmax} \left( \boldsymbol{Mask}(t_1,t_2)_j \frac{\boldsymbol{Q}_i \cdot \boldsymbol{K}_j^T}{\sqrt{d_k}} \right)
 \cdot \boldsymbol{V}_j.
\end{equation}
Here, $\boldsymbol{emb}^*(t_1, t_2)$ is the final representation of the tuple pair $(t_1, t_2)$, and $d_k$ represents the size of each attribute embedding. $\boldsymbol{Mask}(t_1,t_2)$ is the concatenation of $\boldsymbol{Mask}(t_1)$ and $\boldsymbol{Mask}(t_2)$, which is used to mask the impact of a missing attribute value relative to other attributes. 
$\boldsymbol{Q},\boldsymbol{K}, \boldsymbol{V}$ are the query matrix, the key matrix, and the value matrix, respectively, which are computed using a linear transformation~\cite{vaswani2017attention}.

\par Finally, $\boldsymbol{emb}^*(t_1, t_2)$ is the input, and we use an MLP to output a binary decision $y$, indicating whether $t_1$ and $t_2$ can be integrable. 

\subsection{Adversarial Trainer}~\label{sec: trainer}

The \trainer serves dual purposes: (1) establish a contrastive training objective optimized using an SGD-based algorithm; (2) identify adversarial examples to further enrich the training set. 
Next, we will explain this idea in detail.

\noindent\textbf{Objective Function.} In this work, we use a binary noise contrastive estimation (NCE) loss function~\cite{nips_scl_khosla2020supervised} for the training objective, which is formulated as:
\begin{equation}\label{eq: objective}
\mathcal{L} = \sum_{i=1}^{N} (\sum_{j=1}^{N_{pos}} log~f(t_i, t^+_{ij}) + \sum_{j=1}^{N_{neg}} log~f(t_i, t^-_{ij})).
\end{equation}

Here, $N$ is the total number of tuples in the training table, and $t_{ij}^+$ and $t^-_{ik}$ denote one positive instance and negative instance for the tuple $t_i$, respectively.
The optimization of the NCE loss function enables the model to make similar tuples that are close in the embedding space while scattering dissimilar tuples.

\noindent\textbf{Adversarial Examples and Training.}  
We propose the use of adversarial examples~\cite{ilyasnlps2019adversarial,zhangtnnls2019adversarial} to further enrich the collection of positive training instances.
Technically speaking, an adversarial example is typically viewed as a perturbed version of the original input example, which results in a significant impact in a decision made by a machine learning model.
Perturbations in data augmentation that operate on the original tuples while perturbations in adversarial training directly impact the embedding vectors, expressed as:
\begin{equation}
p(\boldsymbol{emb}(t_i)) = \boldsymbol{emb}(t_i) + \boldsymbol{r},
\end{equation}
where $\boldsymbol{r}$ denotes the perturbation vector. 
Since the adversarial example has the most significant impact on model performance, this can also be expressed as the following objective function:
\begin{equation}\label{eq：adversarial examples}
max_{\boldsymbol{r}} \space \mathcal{L}(f(t_i, t_i^{adv}),y_i), s.t.\space ||\boldsymbol{r}||_2 < \epsilon.    
\end{equation}

Here, $\epsilon$ is a small value that constrains the magnitude of the perturbation, and $y_i$ is set to 1 because the pair consisting of the original tuple and its corresponding adversarial example, $(t_i, t^{\text{adv}}_i)$, should consistently yield a positive prediction.

\par Since $\boldsymbol{r}$ is very small, the loss function is approximately equivalent to the following equation derived using a first-order Taylor approximation~\cite{NoceWrig06}:
\begin{equation}\label{eq: taylor approximation}
    \mathcal{L}(f((t_i, t_i^{adv}),y_i) \approx  \mathcal{L}(f((t_i, t_i),y_i) + \nabla_{t_i} \mathcal{L}(f((t_i, t_i),y_i)^T\boldsymbol{r}.
\end{equation}

\par When using a Lagrange Multiplier Method~\cite{lagrange-multiplier-book} to solve Eq.~\ref{eq：adversarial examples} and Eq.~\ref{eq: taylor approximation}, we get:
\begin{equation}
    \boldsymbol{r}=-\epsilon\frac{\nabla_{t_i}\mathcal{L}(f((t_i, t_i),y_i)}{||\nabla_{t_i}\mathcal{L}(f((t_i, t_i),y_i)||^2}
\end{equation}

\par Consequently, for each tuple $t$, we can derive an adversarial example $t_i^{adv}$.
This process is repeated in each training epoch, with $(t_i,t_i^{adv})$ being added to the positive training instance pool.
\par Once we have trained the model $f$, it can be employed to assess the pairwise integrability of any tuple pair in the table $T$. 
Similar to entity resolution, this process can be accelerated using blocking techniques~\cite{deeper_ebraheem2018distributed,thirumuruganathan2021deepbl}. These techniques partition tuples into distinct blocks, enabling pairwise comparisons exclusively within each block, thereby enhancing computational efficiency. In this work, we adopt the LSH-based blocking approach proposed in~\cite{deeper_ebraheem2018distributed} as the default method. Additionally, the blocking technique reduces the computational burden of \taskb, as integrable sets need to be identified only within a single block.

\section{\TASKB}
\label{sec: integrable set identification}
Given the integrability of any two tuples in Table $T$, determined using a binary classifier, this section introduces how to derive all possible integrable sets within $T$.
Most existing studies, such as ALITE~\cite{alite_khatiwada2022integrating}, are limited to integrating two tuples in each step of the algorithm.
This restricts their ability to include additional tuples that could potentially resolve conflicts within larger sets.
In contrast, \emph{our objective is to integrate multiple tuples simultaneously}, enabling the detection and resolution of conflicts as they arise. 
Therefore, our goal is to identify all \add{integrable sets} within Table $T$, where each integrable set refers to a set of tuples that can be integrated, as defined in Def.~\ref{def: integrable set discovery}. As discussed in Sec.~\ref{def: integrable set discovery}, each integrable set found in table $T$ should meet the criteria of consistency and maximality. To achieve this, we can frame the task of \taskb as the problem of identifying all maximal cliques in a graph, as described below.

\noindent \textbf{Integrable Set Discovery by Finding Maximal Cliques}. To discover integrable sets within each table $T$, we proceed by constructing an undirected graph $\mathcal{G}=\{\mathcal{V}, \mathcal{E}\}$ from $T$ in two stages: 
1) each tuple $t_i$ is represented as a node $v_i \in \mathcal{V}$; 2) for every tuple pair $(t_i, t_j)$, if the binary classifier $f(t_i,t_j)=1$, an edge $e_{ij} \in \mathcal{E}$ is created. 
Once $\mathcal{G}$ is constructed, the task of \taskb reduces to the task of finding all maximal cliques in $\mathcal{G}$. 
In graph theory, a clique is a subset of vertices in an undirected graph such that every pair of vertices in this subset is connected by an edge, and a maximal clique in a graph is a clique that cannot be extended by adding another adjacent vertex from the graph.
Thus, there exists a one-to-one mapping for each integrable set in table $T$ to each maximal clique in graph $\mathcal{G}$. 

This allows us to leverage graph algorithms designed for clique detection to efficiently find all possible integrable sets within $T$.
One of the most common algorithms for this problem is the Bron-Kerbosch algorithm~\cite{regneri2007bronalg}, which uses recursive backtracking and a pivot selection strategy to efficiently explore and prune the search space.
Algorithm~\ref{alg:bronkerbosch} outlines the key steps of the Bron-Kerbosch algorithm, in which three sets $R$, $P$, and $X$ are used: $R$ denotes the current clique being constructed; $P$ denotes the set of the candidate vertices that can potentially be added to $R$; $X$ denotes the set of vertices that have already been excluded from consideration. 
The algorithm proceeds as follows: 
If both $P$ and $X$ are empty, then $R$ is a maximal clique, and the algorithm will output $R$ (Lines 2-4). 
Then, we select a pivot vertex $v_i$ from $P \cup X$ (Line 5). 
For each vertex $v_j \in P$ that is not adjacent to $v_i$, we add $v_j$ to $R$ and intersect both $P$ and $X$ with the neighbors of $v_j$ (Lines 6-7). 
After exploring all vertices in $P$ that are not adjacent to $v_i$, each vertex $v$ is moved to $X$ and the algorithm continues (Lines 8-9). 

\begin{algorithm}
\SetAlgoLined
\SetKwFunction{BronKerbosch}{BronKerbosch}
\SetKwProg{Fn}{Function}{:}{}

\Fn{\BronKerbosch{$R, P, X$}}{
    \If{$P$ is empty and $X$ is empty}{
        \textbf{Output:} $R$\;
    }
    $u \leftarrow$ ChoosePivot($P \cup X$)\;
    \ForEach{$v \in P \setminus N(u)$}{
        \BronKerbosch{$R \cup \{v\}, P \cap N(v), X \cap N(v)$}\;
        $P \leftarrow P \setminus \{v\}$\;
        $X \leftarrow X \cup \{v\}$\;
    }
}

\caption{The Bron-Kerbosch Algorithm}
\label{alg:bronkerbosch}
\end{algorithm}

\noindent \textbf{Integrable Set Discovery by Community Detection.}
In theory, Algorithm~\ref{alg:bronkerbosch} can accurately identify all integrable sets from table $T$ if the integrability of any tuple pair in $T$ can be correctly predicted.
However, Algorithm~\ref{alg:bronkerbosch} is not empirically robust for our problem. 
Consider an integrable set $S$ exists and a tuple $t_i \in S$ should be integrable with any other tuple $t_j \in S$. 
If \methodone incorrectly determines the integrability of $t_i$ and any other tuple $t_j \in S$, Algorithm~\ref{alg:bronkerbosch} may fail to add $t_i$ into $S$, even if the integrability of $t_i$ and other tuples $t_k \in S \land t_k \neq t_j$ can be correctly decided. 
To address this issue, we relax the topographical requirement of an integrable set in graph $\mathcal{G}$ in practice, as described in Def.~\ref{def: relaxed integrable set discovery}.
Instead of strictly requiring that each tuple pair in an integrable set is judged as integrable by \methodone, we only ensure that each tuple within an integrable set is integrable with the majority of tuples in the same set. 
This approach transforms the task of \taskc into a community detection problem, which aims to identify densely interconnected groups or clusters of nodes within the graph. 
In this work, we have selected a collection of representative community detection methods and applied them to the task of \taskb, as detailed below:
\begin{itemize}[noitemsep, leftmargin=*]
\item The \textbf{Louvain~\cite{louvain}} algorithm is a modularity-based method that iteratively optimizes the modularity to find a partition of the network that maximizes the quality of the community structure.
\item \textbf{Newman-Girvan~\cite{newman2004NGalg}} algorithm is a hierarchical clustering method that hierarchically removes edges with ``high betweenness centrality'' to identify communities.
\item \textbf{Infomap~\cite{rosvall2008infomap}} is an information-theoretic approach that minimizes the description length of a random walk path through the network to uncover communities.
\item \textbf{Spectral Clustering~\cite{ng2001spectral}} uses eigenvectors from a graph Laplacian matrix to partition nodes into clusters.
\item \textbf{Graph Neural Network (GNN)~\cite{bruna2017community}} is a DL-based representative learning approach that aims to leverage graph neural networks to learn meaningful representations of nodes based on the topographical structure and attributes, which are then used to cluster nodes into different communities.
\end{itemize}

As detailed in Sec.~\ref{sec: experiment}, we will evaluate the effectiveness and efficiency of each method mentioned above in addressing the task of \taskb.

\section{\TASKC}\label{sec: multi-tuple integration}

\begin{figure*}
    \centering
    \includegraphics[width=\linewidth]{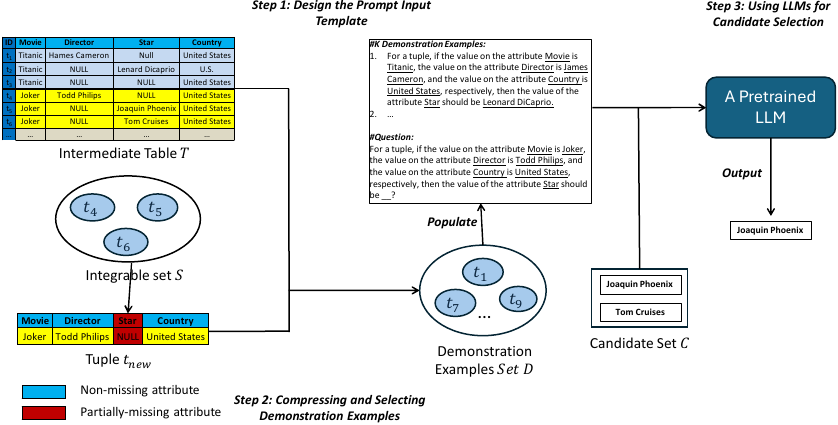}
    \caption{The workflow of \methodtwo}
    \label{fig: iclcf}
\end{figure*}
\subsection{The Key Idea}

Based on \taska, we obtain a collection of integrable sets $\mathcal{S}$.
For each integrable set $S\in\mathcal{S}$, we integrate all of the tuples contained in $S$ into a \emph{single} tuple $t_{new}$, which provides more comprehensive information than each individual tuple $t\in S$.
Thus, the question now is how to determine the correct attribute value $t_{new}[a]$ for an attribute $a$, which can fall into one of three potential categories: 
\begin{itemize}[noitemsep, leftmargin=*]
\item \textbf{Missing-value Attributes.} An attribute $a$ is considered as a missing value attribute if $t[a] == NULL$ for all $t \in S$.
In this case, $t_{new}[a]$ is assigned a $NULL$ value.
This choice is made due to the absence of any reliable information to populate $t_{new}[a]$.
\item \textbf{Unique-value Attributes.} An attribute $a$ is considered unique attribute if there is only one valid value $v$ in $a$. In this case, $v$ is naturally selected as the correct value for $a$.
\item \textbf{Multiple-value Attribute.} An attribute $a$ is considered to have multiple values if at least one tuple contains more than one value for $a$, which may conflict with each other. This results in a candidate set $C(a) = \{t[a] \mid t \in S \land t[a] \neq \text{NULL}\}$.  
To resolve this, the correct value $v^* \in C(a)$ must be selected to fill $t_{\text{new}}[a]$.
\end{itemize}

\par The first two types of attributes can be resolved easily, and hence we focus on how to fill multiple-value attributes.
This can be formulated as a conflict resolution problem, as articulated in Def.~\ref{def: conflict resolution}.
Existing solutions for conflict resolution in the context of a data fusion problem rely on truth discovery approaches~\cite{voting_li2012truth,vldb_df_li2014confidence,sigkdd_df_zhi2015modeling}, which first estimate the trustworthiness of a data source, and subsequently choose the value from the most reliable source to fill a missing value.
However, truth discovery approaches typically rely on a lot of training data or metadata (i.e., paper citation and reviews) in order to estimate the reliability of the source.
This imposes limitations on the applicability of these methods, as practical scenarios often lack access to sufficient labeled data, and obtaining such data requires considerable human curation and financial burdens.

\par To address this problem, we propose a novel method for conflict resolution, namely In-context Learning for Conflict Resolution (\methodtwo). This method draws inspiration from the recent success of in-context learning (ICL) within the field of natural language processing~\cite{min2022rethinkingicl,xie2021explanationicl,akyurek2022learningicl} . 
ICL-based methods solve downstream tasks by leveraging contextual information provided by an input prompt, removing the necessity for explicit task-specific training. 
This approach allows large language models (LLMs) to dynamically adapt their behavior based on a few labeled demonstration examples and instructions embedded in the prompt. 
Intuitively, employing ICL-based methods for conflict resolution effectively addresses the issue of insufficient labeled data. These methods leverage LLMs that are pre-trained on extensive corpora, thereby incorporating a broad base of general knowledge. Furthermore, in private or specialized domains not extensively covered by the training corpora, ICL-based methods require only a few examples to adapt and deliver robust performance.

However, applying ICL-based methods to our conflict resolution problem presents several challenges.
Technically, the effectiveness of these methods heavily relies on the selection of demonstration examples.
While ICL-based methods typically require only a small number of demonstration examples, including more examples may provide LLMs with richer contextual information to handle downstream tasks effectively.
However, the constrained input size of current LLMs limits the number of demonstration examples that can be incorporated. 
Thus, a key challenge is to maximize the number of demonstration examples for a given input size.
Furthermore, given the restricted number of demonstration examples that can be selected, it is crucial to prioritize the ones that are most relevant to the specific aspects of the downstream task.
This ensures that the model can effectively generalize to the provided examples, thereby enhancing the ability to resolve conflicts accurately.

To address the above challenges in leveraging ICL-based methods  for conflict resolution, \methodtwo relies on two key strategies: demonstration example compression and selection. 
Specifically, in demonstration example compression, non-relevant attributes that contribute minimal information to the prediction of target attributes are ignored when transforming a tuple into a natural language sentence. 
This reduces the number of tokens to express each demonstration example. 
Furthermore, in demonstration example selection, we judiciously choose the demonstration examples that are semantically similar to the target, ensuring that the most relevant examples are included.

 Fig.~\ref{fig: iclcf} illustrates the workflow of our proposed \methodtwo, which mainly consists of three steps that apply in-context learning to solve the conflict resolution problem to integrate tuples in each integrable set:
\begin{itemize}[leftmargin=*]
\item \textit{Step 1: Designing the Prompt Input} — This step reformulates the conflict resolution task into a prompt format suitable for resolution using an LLM.
\item \textit{Step 2: Compressing and Selecting Demonstration Examples} — This step selects the most representative and informative tuples from the table 
$T$ to serve as demonstration examples.
\item \textit{Step 3: Using LLMs for Candidate Selection} — This step employs an LLM to predict which value in the candidate set is most likely correct, based on the provided prompt input.
\end{itemize}

\subsection{Designing the Prompt Input Template}
The first step is to transform the problem of conflict resolution into a prompt input (or context) $P$ with several demonstration examples, which can be formally expressed as $P=\{x_1,y_1,x_2$, $y_2,...,x_k,y_k,\hat{x},\hat{y}\}$, where $x_i$ and $y_i$ denotes an answer and a question in the form of natural language sentence, respectively, jointly constituting a demonstration example $d_i$. $\hat{x}$ is the target question and $\hat{y}$ is the undecided answer. 

In this problem, for an integrable set $S$, to fill a multiple-value attribute $a_{mul}$ in $t_{new}$, at most $k$ tuples are selected as demonstration examples to instruct LLMs to make prediction for $t_{new}[a_{mul}]$.
To enable LLMs to understand the demonstrate example, we need to transform each selected tuple $t$ into a natural language sentence composed of $x_i$ and $y_i$.
To accomplish this, for each demonstration tuple $t_i$, we adopt the following template to transform it into a sentence:

\smallskip
\noindent \textbf{Demonstration Example Template}. For a tuple, if the values of an attribute \_\_ is \_\_ , the attribute \_\_ is \_\_, 
..., respectively, then the value of the attribute \_\_ should be \_\_. 
\smallskip

In more detail, to transform a tuple $t_i$ into a demonstration example $d_i$ in terms of making prediction on $a_{mul}$ in an integrable set $S$, each of its non-missing attribute names (except for the multiple-value attribute $a_{mul}$ in $S$) and the corresponding values should be used to fill the blank in the ``if" clause, forming $x_i$, and the name of the partially attribute $a_{mul}$ and the value on $a_{mul}$ of $t_i$ are used to fill the blank in the ``then" clause, forming $y_i$.
Similarly, each of its non-missing attribute names of $a_{new}$ and the corresponding values should be used to fill the blank in the ``if" clause, forming $\hat{x}$, while the blank for the attribute value in the ``then" clause should be left out, which will be answered by an LLM.
Thus, all demonstration examples along with the target question constitute the prompt input $P$.
Next, we further illustrate how to judiciously select demonstration examples to populate the input prompt $P$.

\subsection{Compressing and Selecting Demonstration Examples}
Demonstration examples are labeled data that instruct LLMs how to make predictions for a given task. 
The choice of demonstration examples, denoted as $D$, has a significant impact on the effectiveness of in-context learning~\cite{brinkmann2023iclexample}.
Intuitively, the model performance of \methodtwo is closely relevant to two factors: 1) the number of demonstration examples $N$ in the prompt input $P$; 2) the relevance of the demonstration examples in $D$ to the target question $q$ transformed from $t_{new}$. To address these two issues, in this paper, we propose effective demonstration example compression and selection strategies in Section \ref{sec: compression} and Section \ref{sec: selection}, respectively.

\subsubsection{Demonstration Example Compression}~\label{sec: compression}
Since the input size of an LLM is limited, it is impossible to include as many demonstration examples in the prompt input $P$ as desired.
Thus, a key challenge here is how to include as many demonstration examples in the prompt input $P$ as possible.
To address this problem, we propose a demonstration example compression strategy.
Intuitively, if we want to include as many demonstration examples in $P$ as possible, one solution is to minimize the average number of tokens to express a demonstration example in the natural language sentence while maintaining the information about the target attribute.
When we transform a tuple $t_i$ into a demonstration example $d_i$, the length of $d_i$ is roughly proportional to the number of the non-missing attributes in $t_i$. However, not all non-missing attributes contribute to the prediction of $a_{mul}$. 
For example, the ID number is not relevant and can be ignored if we want to make a prediction on the position of a person. 
Thus, if we can remove all the non-relevant attributes for $a_{mul}$, the average length of demonstration examples can be greatly reduced, and more demonstration examples can be included in the prompt input $P$ to instruct LLMs to achieve better performance. 
Note that even in the case where the maximum number of demonstration examples that can be included in the prompt input is larger than the number of available labeled demonstration examples, the proposed demonstration example compression strategy is still helpful because it reduces the length of prompt input, and the inference time of LLMs can be reduced, improving the overall efficiency.

Then, to quickly and effectively decide whether a non-missing attribute $a_{non}$ is not relevant when predicting $a_{mul}$, the mutual information (MI) metric can be used, which is widely used to capture the relationship and dependency between two variables.
Formally, given two variables $X$ and $Y$, their mutual information is calculated using 
\begin{equation}~\label{eq: mi}
I(X; Y) = \sum_{x \in X} \sum_{y \in Y} p(x, y) \log \left( \frac{p(x, y)}{p(x) p(y)} \right),
\end{equation}
where $p(x,y)$ denotes the joint distribution possibility for $X=x$ and $Y=y$, and $p(x)$ and $p(y)$ denote the marginal distribution for $X$ and $Y$, respectively.
Specifically, a large $I(X;Y)$ indicates that there is a strong dependence between $X$ and $Y$ and when $I(X;Y)$ is close to 0, it means that there is no dependency between the two variables.

\par In this paper, we utilize Eq.~\ref{eq: mi} to calculate the dependency between a non-missing attribute $a_{non}$ and a multiple-value attribute $a_{mul}$. 
Since Eq.~\ref{eq: mi} only applies to discrete variables, we process different attribute types as follows: 
1) \textit{Categorical and textual attributes} are transformed into discrete values based on their distinct values; 
2) \textit{Numerical attributes} are converted into discrete values by binning them into intervals. 
Then, we apply a small value $\beta=0.1$ as the threshold to decide if the two attributes are dependent.
Specifically, if $I(a_{mul};a_{non})<\beta$, we remove the corresponding description for the non-missing attribute $a_{non}$ in the demonstration example template.

\subsubsection{Demonstration Example Selection}~\label{sec: selection}
The choice of demonstration examples also provide important performance improvements for the LLMs during conflict resolution.
It is desirable to choose the tuples that are most relevant to the new tuple $t_{new}$ as demonstration examples. 
In this paper, we investigate three different demonstration selection strategies, namely a random selection strategy, a $K$-NN selection strategy, and a weighted $k$-NN selection strategy, respectively.

\noindent\textbf{Random Selection.} An intuitive and simple strategy is to randomly choose $k$ tuples from the table $T$ as demonstration examples. 
However, since the demonstration examples are randomly selected from $T$, it is likely that they contain limited information to help make a prediction for $a_{mul}$ and $t_{new}$.

\noindent\textbf{$k$-NN Selection.} To address problem faced by random selection above, we select tuples that are semantically close to $t_{new}$ as demonstration examples. 
To accomplish this goal, we compute the cosine similarity using an embedding-based representations of a candidate tuple $t$ and the new tuple $t_{new}$.
Specifically, for each integrable set $S \in \mathcal{S}$, an initial $t_{new}$ is derived by collating all available values from non-null attributes within the integrable set, which produces a set of multiple $t_{new}$ of size $|\mathcal{S}|$.
Furthermore, we encode each $t_{new}$ using Eq.~\ref{Eq: tuple embedding} to generate an encoded representation $\boldsymbol{emb}(t_{new})$, and then compute the average embedding $\overline{\boldsymbol{emb}(t_{new})}$ for all $\boldsymbol{emb}(t_{new})$ of size $|\mathcal{S}|$.
Next, we perform a $k$-NN search to identify the $k$ tuples from table $T$ that are the most similar to $\overline{\boldsymbol{emb}(t_{new})}$ to the demonstration examples. 

\noindent\textbf{Weighted $k$-NN Selection.} While the $k$-NN selection strategy is able to select relevant demonstration examples, as described in Sec.~\ref{sec: compression}, different attributes in a candidate tuple $t$ can have different degrees of impact when predicting $a_{mul}$, which cannot be determined using $k$-NN selection alone. 
To this end, we further improve $k$-NN selection by weighting different attributes using their normalized mutual information with the non-missing attribute $a_{mul}$ when we aggregate the attribute representations into the tuple representation.

\subsection{Using LLMs for Candidate Selection}
Finally, the step of candidate selection focuses on using LLMs to select the correct value for attributes where multiple conflicting values exist.
With the design of the prompt input template and the selected demonstration examples, we can obtain the entire prompt input $P$, and ask an LLM to answer the question -- filling in $\hat{y}$. 
Here, the set of candidate values $C$ for $\hat{y}$ is composed of tuples whose values on $a_{mul}$ are not missing. 
Then, an LLM is used to predict which candidate value $v \in C$ has the highest likelihood of appearing in the prompt input $P$. 
This can be defined as $max_{v \in C} P_{LLM}(\hat{y}=v|P)$. 
In Sec.~\ref{sec: experiment}, we investigate the performance of various LLMs for the conflict resolution task.

\section{Data Preparation and Evaluation}\label{sec: data preparation}
This section describes how to create the benchmarks, followed by the evaluation framework used in our experiments. 

\subsection{Benchmark Creation}\label{sec: dataset}
An ideal benchmark must exhibit two critical characteristics: 
(1) every dataset should contain semantic equivalence cases, typographical errors, and conflicts, all of which we aim to address; 
(2) a dataset should be accompanied by definitive ground-truth annotations that label integrable sets and correct values for conflict resolution, in order to facilitate a thorough effectiveness evaluation of our proposed solution.
Given the presence of the first characteristics, directly using the benchmarks proposed for ALITE~\cite{alite_khatiwada2022integrating}, the most closely related work to ours, is not sufficient since errors and conflicts are not supported in ALITE.
Thus, we have created our own benchmarks by injecting semantic equivalences, typographical errors, and conflicts, and recording ground-truth annotations for the evaluation. 
In Sec.~\ref{sec: further discussion}, we also discuss why benchmarks from related tasks, such as entity resolution and conflict resolution, are not suitable for our problem

We create our benchmarks using two dataset repositories from ALITE~\cite{alite_khatiwada2022integrating}, \textbf{Real} and \textbf{Join}~\cite{alite_khatiwada2022integrating}.
Both dataset repositories contain multiple datasets, each of which contains a set of input tables to be integrated.

Furthermore, we perform an outer-union operator~\cite{alite_khatiwada2022integrating,bleiholder2010subsumption} to join the tables in each dataset and produce a single intermediate table $T$, which is used to create our benchmarks. 
For each dataset repository, all the datasets are divided into three categories according to the size of the number of intermediate tables $T$, namely \emph{Small}, \emph{Medium}, and \emph{Large}.
Table~\ref{table: statistics} provides additional statistics of the two dataset repositories used in the benchmark. Furthermore, in this paper, Rn refers to a specific dataset in the Real repository, and Jm refers to a specific dataset in the Join repository.

\begin{table*}[tp]
\captionsetup{skip=3pt}
\centering
\footnotesize
\caption{The statistics of the two dataset repositories, where $\alpha_1$ and $\alpha_2$ denotes the ratio of numerical attributes and missing values}\label{table: statistics}
\begin{tabular}{|cc|c|c|c|c|c|c|c|}
\hline
\multicolumn{2}{|c|}{\multirow{2}{*}{}}                                & \multirow{2}{*}{\textbf{\#Datasets}} & \textbf{\#Average} & \textbf{\#Average} & \textbf{\#Average} & \textbf{\#Average} & \multirow{2}{*}{\textbf{$\alpha_1$}} & \multirow{2}{*}{\textbf{$\alpha_2$}} \\
\multicolumn{2}{|c|}{}                                                 &                                      & \textbf{Tables}    & \textbf{Columns}   & \textbf{Rows}      & \textbf{Clusters}  &                                      &                                      \\ \hline
\multicolumn{1}{|c|}{\multirow{3}{*}{\textbf{Real}}} & \textbf{Small}  & 4                                    & 9.5                & 27                 & 1524.2             & 194.9              & 10\%-25\%                            & 0\%-25\%                             \\ \cline{2-9} 
\multicolumn{1}{|c|}{}                               & \textbf{Medium} & 4                                    & 9                  & 18                 & 6902.5             & 470.6              & 0\%-10\%                             & 20\%-55\%                            \\ \cline{2-9} 
\multicolumn{1}{|c|}{}                               & \textbf{Large}  & 3                                    & 9.3                & 12                 & 48617.5            & 1372.3             & 5\%-10\%                             & 35\%-45\%                            \\ \hline
\multicolumn{1}{|c|}{\multirow{3}{*}{\textbf{Join}}} & \textbf{Small}  & 9                                    & 5                  & 9.2                & 4578.2             & 378.3              & 15\%-25\%                            & 15\%-20\%                            \\ \cline{2-9} 
\multicolumn{1}{|c|}{}                               & \textbf{Medium} & 9                                    & 12.5               & 8.5                & 33704.4            & 1124.7             & 5\%-20\%                             & 25\%-35\%                            \\ \cline{2-9} 
\multicolumn{1}{|c|}{}                               & \textbf{Large}  & 10                                   & 15.8               & 13.7               & 77351.2            & 2617.3             & 15\%-20\%                            & 10\%-20\%                            \\ \hline
\end{tabular}
\end{table*}

\subsubsection{Noise Injection}
\par First, we inject noise into the clean table $T$, so as to simulate semantic equivalence and typographical errors using a pre-define error rate.
Since the datasets used in the literature for data cleaning~\cite{abedjan2016detecting,peng2022garf} have a noise rate between 5\% and 40\%, we set the default noise rate to 30\% in our experiments.
Furthermore, we test three different settings for the ratio between semantic equivalence and typographical errors, 10\%$\slash$20\% (SE-heavy, short for semantic equivalence-heavy), 15\%$\slash$15\% (Balanced), and 20\%$\slash$10\% (TE-heavy, short for typographical error-heavy).
Unless specified otherwise, we use the balanced noise case by default.

\begin{itemize}[noitemsep, leftmargin=*]
\item \textbf{Semantic Equivalence.} Backtranslation~\cite{edunov2018understanding,shorten2021text} has been widely employed to generate sentences that maintain similar semantic meaning to the original sentence in the field of natural language processing. 
Once a cell is chosen for noise injection, we use backtranslation~\cite{edunov2018understanding,shorten2021text} to generate an alternative description with similar meaning to replace the original value.
We only inject this type of noise in textual attributes. 
\item \textbf{Typographical Errors.} We simulate typographical errors in a comprehensive way, such as character swapping and character deletion.
More details about this setting can be seen in the technical report~\cite{github}.
We inject typographical errors for both textual and numerical attributes. 
\end{itemize}

\subsubsection{Conflict Generation}
To generate conflicts, for each integrable set in $T$ (we will introduce how to obtain the ground-truth integrable set below), we choose non-missing textual attribute $a$ (introduced in Sec.~\ref{sec: multi-tuple integration}), and a tuple $t$ that has a non-null value for the attribute $a$.
We use tuple $t$ as a template to generate the conflict tuples.
Specifically, we assume that the conflict tuples come from different resources, each of which has a reliability score $r$ between 0 and 1.
Then, we randomly replace $t[a]$ with another value from the other values in the attribute $a$ with a probability $1-r$.
Using this strategy, we create 3-5 conflict tuples for each integrable set.
We set $r_i$ to a random value between 30\%-80\%.

\subsubsection{Ground-truth}\label{sec: ground-truth}
We also need to create the ground-truth for our benchmarks.
For multi-tuple integration, we use ALITE~\cite{alite_khatiwada2022integrating} to integrate tuples in an error-free and conflict-free manner.
During the integration, we track tuple pairs that are integrated, which are considered to be part of the ground-truth.
Then based on the results from from ALITE, we use the Bron-Kerbosch algorithm to find the integrable sets in $T$. 
Note that the Bron-Kerbosch algorithm produces correct results in this case, as the input of pairwise integrability is error-free.
Thus, in our benchmarks, the ground-truth can be generated by obtaining the tuples that form the integrable sets.
For conflict resolution, the template tuple value $t[a]$ for the attribute $a$ is the ground-truth since we create the conflict tuples based on the tuple $t$.

\subsection{Evaluation Metrics}\label{sec: evaluation metrics}
\myparagraph{Metrics for The Task of~\TASKA}
\Taska serves a similar purpose to the entity resolution task~\cite{deeper_ebraheem2018distributed,ditto_li2020deep,deepmatcher_mudgal2018deep}, so we use \emph{F1} to evaluate our results, and is expressed as the harmonic mean of $Recall$ and $Precision$, denoted as $F_1 = 2 \times \frac{Recall \times Precision} {Recall + Precision}$.
Here, $Recall = \frac{T_P}{T_P+F_N}, Precision = \frac{T_P}{T_P+F_P}$, where $T_P$ denotes the number of integrable tuple pairs that are determined to be integrable, $F_N$ represents the number of integrable tuple pairs that are incorrectly determined to be non-integrable, and $F_P$ indicates the number of tuple pairs that are not integrable but determined to be integrable.

\myparagraph{Metric for The Task of \TASKB}
Given a set of ground-truth integrable sets $\mathcal{R} = \{R_1, R_2, ..., R_n\}$ and a set of integrable sets $\mathcal{S} = \{S_1,S_2$, $...,S_m\}$, we use the \emph{Similarity} between $\mathcal{R}$ and $\mathcal{S}$ as the evaluation metric.
\emph{Similarity} is defined as the maximum weighted matching score in a bipartite graph, which has two disjoint sets of vertices $\mathcal{R}$ and $\mathcal{S}$, and a set of edges $(R_i,S_j)$ between any two vertices from $\mathcal{R}$ ($R_i \in \mathcal{R}$) and $\mathcal{S}$ ($S_j \in \mathcal{S}$), respectively.
The weight of each edge $(R_i,S_j)$ is computed using Jaccard similarity, where $J=\frac{|R_i\cap S_j|}{|R_i\cup S_j|}$.

\myparagraph{Metric for The Task of~\TASKC} 
We evaluate several methods in terms of \emph{Accuracy}, which is the ratio computed by dividing the number of correctly filled attributes by the total number of conflict attributes available.

\subsection{Further Discussion on Other Benchmarks}~\label{sec: further discussion}
Although previous work~\cite{voting_li2012truth,deepmatcher_mudgal2018deep,deeper_ebraheem2018distributed} on entity resolution and conflict resolution provide test collections, they cannot be used in our experiments for the following reasons:
\begin{itemize}[leftmargin=*,noitemsep]
\item \textbf{Entity Resolution}. Most entity resolution benchmarks~\cite{deeper_ebraheem2018distributed,deepmatcher_mudgal2018deep} focus on matching pairs of tuples, restricting integrable sets to a size of two and eliminating the need for conflict resolution. Furthermore, these benchmarks typically feature few missing values and a limited number of erroneous data, e.g., typographical errors, which we aim to address.
\item \textbf{Conflict Resolution}. In most conflict resolution benchmarks~\cite{vldb_df_li2014confidence}, conflicts are not considered in the context of table integration, and the size of integrable sets, in terms of the number of tuples (items), is relatively small. 
\end{itemize}

\section{Experiments}\label{sec: experiment}

First, we evaluate the effectiveness of the proposed methods for the three core tasks studied in this work: \emph{\taska}, \emph{\taskb}, and \emph{\taskc} (Sec.~\ref{sec: effectiveness study}). 
Second, considering that the proposed \methodone and \methodtwo are designed to be effective using limited labeled data, we assess the performance under this constraint (Sec.~\ref{sec: label-efficiency study}).
Third, we investigate the impact of various choices of PLMs and LLMs on \methodone and \methodtwo, respectively, considering they play important roles in determining the quality of feature representations and the overall performance of the methods in diverse scenarios. (Sec.\ref{sec: LLM Choice Study}).
Fourth, we conduct an ablation study and a hyper-parameter study to thoroughly analyze the behavior of \methodone and \methodtwo (Sec.~\ref{sec: hyper-parameter study} and Sec.~\ref{sec: ablation study}).

\subsection{Experimental Setup}

\subsubsection{Environment}
We implement all the algorithms in Python 3.9 and run the experiments on an Ubuntu sever equipped with an Intel(R) Core(R)  13700KF CPU and RTX 4090 GPU.
The source code is available at~\cite{github}. 

\begin{table*}[tp]
\centering
\caption{\emph{F1} and \emph{Similarity} using balanced noise}~\label{table: balanced}
\begin{tabular}{|c|c|cccc|cccc|}
\hline
\multirow{2}{*}{\textbf{Methods}}  & \multirow{2}{*}{\textbf{Metrics}} & \multicolumn{4}{c|}{\textbf{Real}}                                                                                                  & \multicolumn{4}{c|}{\textbf{Join}}                                                                                                  \\ \cline{3-10} 
                                   &                                   & \multicolumn{1}{c|}{\textbf{Overall}} & \multicolumn{1}{c|}{\textbf{Small}} & \multicolumn{1}{c|}{\textbf{Medium}} & \textbf{Large} & \multicolumn{1}{c|}{\textbf{Overall}} & \multicolumn{1}{c|}{\textbf{Small}} & \multicolumn{1}{c|}{\textbf{Medium}} & \textbf{Large} \\ \hline
\multirow{2}{*}{\textbf{ALITE}}    & \emph{F1}                         & \multicolumn{1}{c|}{0.209}            & \multicolumn{1}{c|}{0.214}          & \multicolumn{1}{c|}{0.198}           & 0.234          & \multicolumn{1}{c|}{0.183}            & \multicolumn{1}{c|}{0.197}          & \multicolumn{1}{c|}{0.174}           & 0.183          \\ \cline{2-10} 
                                   & \emph{Similarity}                 & \multicolumn{1}{c|}{0.146}            & \multicolumn{1}{c|}{0.153}          & \multicolumn{1}{c|}{0.149}           & 0.135          & \multicolumn{1}{c|}{0.125}            & \multicolumn{1}{c|}{0.131}          & \multicolumn{1}{c|}{0.117}           & 0.128          \\ \hline
\multirow{2}{*}{\textbf{Unicorn}}  & \emph{F1}                         & \multicolumn{1}{c|}{0.723}            & \multicolumn{1}{c|}{0.795}          & \multicolumn{1}{c|}{0.705}           & 0.669          & \multicolumn{1}{c|}{0.799}            & \multicolumn{1}{c|}{0.834}          & \multicolumn{1}{c|}{0.806}           & 0.758          \\ \cline{2-10} 
                                   & \emph{Similarity}                 & \multicolumn{1}{c|}{0.568}            & \multicolumn{1}{c|}{0.613}          & \multicolumn{1}{c|}{0.571}           & 0.521          & \multicolumn{1}{c|}{0.641}            & \multicolumn{1}{c|}{0.667}          & \multicolumn{1}{c|}{0.654}           & 0.601          \\ \hline
\multirow{2}{*}{\textbf{SSACL-AE}} & \emph{F1}                         & \multicolumn{1}{c|}{0.741}            & \multicolumn{1}{c|}{0.809}          & \multicolumn{1}{c|}{0.729}           & 0.686          & \multicolumn{1}{c|}{0.801}            & \multicolumn{1}{c|}{0.827}          & \multicolumn{1}{c|}{0.813}           & 0.765          \\ \cline{2-10} 
                                   & \emph{Similarity}                 & \multicolumn{1}{c|}{0.584}            & \multicolumn{1}{c|}{0.617}          & \multicolumn{1}{c|}{0.587}           & 0.550          & \multicolumn{1}{c|}{0.656}            & \multicolumn{1}{c|}{0.671}          & \multicolumn{1}{c|}{\textbf{0.683}}  & 0.616          \\ \hline
\multirow{2}{*}{\textbf{SSACL-DA}} & \emph{F1}                         & \multicolumn{1}{c|}{0.726}            & \multicolumn{1}{c|}{0.801}          & \multicolumn{1}{c|}{0.707}           & 0.664          & \multicolumn{1}{c|}{0.804}            & \multicolumn{1}{c|}{0.837}          & \multicolumn{1}{c|}{0.803}           & 0.760          \\ \cline{2-10} 
                                   & \emph{Similarity}                 & \multicolumn{1}{c|}{0.571}            & \multicolumn{1}{c|}{0.615}          & \multicolumn{1}{c|}{0.569}           & 0.525          & \multicolumn{1}{c|}{0.645}            & \multicolumn{1}{c|}{0.672}          & \multicolumn{1}{c|}{0.652}           & 0.605          \\ \hline
\multirow{2}{*}{\textbf{SSACL}}    & \emph{F1}                         & \multicolumn{1}{c|}{\textbf{0.763}}   & \multicolumn{1}{c|}{\textbf{0.831}} & \multicolumn{1}{c|}{\textbf{0.744}}  & \textbf{0.715} & \multicolumn{1}{c|}{\textbf{0.827}}   & \multicolumn{1}{c|}{\textbf{0.856}} & \multicolumn{1}{c|}{\textbf{0.84}}   & \textbf{0.787} \\ \cline{2-10} 
                                   & \emph{Similarity}                 & \multicolumn{1}{c|}{\textbf{0.610}}   & \multicolumn{1}{c|}{\textbf{0.648}} & \multicolumn{1}{c|}{\textbf{0.608}}  & \textbf{0.574} & \multicolumn{1}{c|}{\textbf{0.670}}   & \multicolumn{1}{c|}{\textbf{0.697}} & \multicolumn{1}{c|}{0.682}           & \textbf{0.632} \\ \hline
\end{tabular}
\end{table*}

\begin{table*}[tp]
\centering
\caption{\emph{F1} and \emph{Similarity} when using SE-heavy noise}~\label{table: SE-heavy}
\begin{tabular}{|c|c|cccc|cccc|}
\hline
\multirow{2}{*}{\textbf{Methods}}  & \multirow{2}{*}{\textbf{Metrics}} & \multicolumn{4}{c|}{\textbf{Real}}                                                                                                  & \multicolumn{4}{c|}{\textbf{Join}}                                                                                                  \\ \cline{3-10} 
                                   &                                   & \multicolumn{1}{c|}{\textbf{Overall}} & \multicolumn{1}{c|}{\textbf{Small}} & \multicolumn{1}{c|}{\textbf{Medium}} & \textbf{Large} & \multicolumn{1}{c|}{\textbf{Overall}} & \multicolumn{1}{c|}{\textbf{Small}} & \multicolumn{1}{c|}{\textbf{Medium}} & \textbf{Large} \\ \hline
\multirow{2}{*}{\textbf{ALITE}}    & \emph{F1}                         & \multicolumn{1}{c|}{0.198}            & \multicolumn{1}{c|}{0.205}          & \multicolumn{1}{c|}{0.201}           & 0.227          & \multicolumn{1}{c|}{0.174}            & \multicolumn{1}{c|}{0,187}          & \multicolumn{1}{c|}{0.169}           & 0.172          \\ \cline{2-10} 
                                   & \emph{Similarity}                 & \multicolumn{1}{c|}{0.134}            & \multicolumn{1}{c|}{0.145}          & \multicolumn{1}{c|}{0.139}           & 0.126          & \multicolumn{1}{c|}{0.124}            & \multicolumn{1}{c|}{0.117}          & \multicolumn{1}{c|}{0.104}           & 0.113          \\ \hline
\multirow{2}{*}{\textbf{Unicorn}}  & \emph{F1}                         & \multicolumn{1}{c|}{0.758}            & \multicolumn{1}{c|}{0.819}          & \multicolumn{1}{c|}{0.739}           & 0.716          & \multicolumn{1}{c|}{0.817}            & \multicolumn{1}{c|}{0.841}          & \multicolumn{1}{c|}{0.829}           & 0.782          \\ \cline{2-10} 
                                   & \emph{Similarity}                 & \multicolumn{1}{c|}{0.594}            & \multicolumn{1}{c|}{0.636}          & \multicolumn{1}{c|}{0.593}           & 0.555          & \multicolumn{1}{c|}{0.657}            & \multicolumn{1}{c|}{0.667}          & \multicolumn{1}{c|}{0.684}           & 0.621          \\ \hline
\multirow{2}{*}{\textbf{SSACL-AE}} & \emph{F1}                         & \multicolumn{1}{c|}{0.764}            & \multicolumn{1}{c|}{0.822}          & \multicolumn{1}{c|}{0.745}           & 0.723          & \multicolumn{1}{c|}{0.824}            & \multicolumn{1}{c|}{0.845}          & \multicolumn{1}{c|}{0.835}           & 0.787          \\ \cline{2-10} 
                                   & \emph{Similarity}                 & \multicolumn{1}{c|}{0.609}            & \multicolumn{1}{c|}{0.648}          & \multicolumn{1}{c|}{0.607}           & 0.574          & \multicolumn{1}{c|}{0.666}            & \multicolumn{1}{c|}{0.678}          & \multicolumn{1}{c|}{0.695}           & 0.626          \\ \hline
\multirow{2}{*}{\textbf{SSACL-DA}} & \emph{F1}                         & \multicolumn{1}{c|}{0.759}            & \multicolumn{1}{c|}{0.822}          & \multicolumn{1}{c|}{0.741}           & 0.715          & \multicolumn{1}{c|}{0.819}            & \multicolumn{1}{c|}{0.842}          & \multicolumn{1}{c|}{0.832}           & 0.783          \\ \cline{2-10} 
                                   & \emph{Similarity}                 & \multicolumn{1}{c|}{0.597}            & \multicolumn{1}{c|}{0.637}          & \multicolumn{1}{c|}{0.595}           & 0.557          & \multicolumn{1}{c|}{0.660}            & \multicolumn{1}{c|}{0.672}          & \multicolumn{1}{c|}{0.686}           & 0.619          \\ \hline
\multirow{2}{*}{\textbf{SSACL}}    & \emph{F1}                         & \multicolumn{1}{c|}{\textbf{0.780}}   & \multicolumn{1}{c|}{\textbf{0.842}} & \multicolumn{1}{c|}{\textbf{0.761}}  & \textbf{0.737} & \multicolumn{1}{c|}{\textbf{0.841}}   & \multicolumn{1}{c|}{\textbf{0.867}} & \multicolumn{1}{c|}{\textbf{0.852}}  & \textbf{0.804} \\ \cline{2-10} 
                                   & \emph{Similarity}                 & \multicolumn{1}{c|}{\textbf{0.627}}   & \multicolumn{1}{c|}{\textbf{0.667}} & \multicolumn{1}{c|}{\textbf{0.624}}  & \textbf{0.591} & \multicolumn{1}{c|}{\textbf{0.687}}   & \multicolumn{1}{c|}{\textbf{0.699}} & \multicolumn{1}{c|}{\textbf{0.714}}  & \textbf{0.650} \\ \hline
\end{tabular}
\end{table*}

\begin{table*}[tp]
\centering
\caption{\emph{F1} and \emph{Similarity} when using TE-heavy noise}~\label{table: TE-heavy}
\begin{tabular}{|c|c|cccc|cccc|}
\hline
\multirow{2}{*}{\textbf{Methods}}  & \multirow{2}{*}{\textbf{Metrics}} & \multicolumn{4}{c|}{\textbf{Real}}                                                                                                  & \multicolumn{4}{c|}{\textbf{Join}}                                                                                                  \\ \cline{3-10} 
                                   &                                   & \multicolumn{1}{c|}{\textbf{Overall}} & \multicolumn{1}{c|}{\textbf{Small}} & \multicolumn{1}{c|}{\textbf{Medium}} & \textbf{Large} & \multicolumn{1}{c|}{\textbf{Overall}} & \multicolumn{1}{c|}{\textbf{Small}} & \multicolumn{1}{c|}{\textbf{Medium}} & \textbf{Large} \\ \hline
\multirow{2}{*}{\textbf{ALITE}}    & \emph{F1}                         & \multicolumn{1}{c|}{0.183}            & \multicolumn{1}{c|}{0.192}          & \multicolumn{1}{c|}{0.187}           & 0.173          & \multicolumn{1}{c|}{0.186}            & \multicolumn{1}{c|}{0.192}          & \multicolumn{1}{c|}{0.187}           & 0.174          \\ \cline{2-10} 
                                   & \emph{Similarity}                 & \multicolumn{1}{c|}{0.145}            & \multicolumn{1}{c|}{0.147}          & \multicolumn{1}{c|}{0.139}           & 0.152          & \multicolumn{1}{c|}{0.132}            & \multicolumn{1}{c|}{0.141}          & \multicolumn{1}{c|}{0.127}           & 0.135          \\ \hline
\multirow{2}{*}{\textbf{Unicorn}}  & \emph{F1}                         & \multicolumn{1}{c|}{0.697}            & \multicolumn{1}{c|}{0.771}          & \multicolumn{1}{c|}{0.681}           & 0.64           & \multicolumn{1}{c|}{0.755}            & \multicolumn{1}{c|}{0.776}          & \multicolumn{1}{c|}{0.773}           & 0.718          \\ \cline{2-10} 
                                   & \emph{Similarity}                 & \multicolumn{1}{c|}{0.562}            & \multicolumn{1}{c|}{0.601}          & \multicolumn{1}{c|}{0.557}           & 0.629          & \multicolumn{1}{c|}{0.617}            & \multicolumn{1}{c|}{0.629}          & \multicolumn{1}{c|}{0.652}           & 0.581          \\ \hline
\multirow{2}{*}{\textbf{SSACL-AE}} & \emph{F1}                         & \multicolumn{1}{c|}{0.717}            & \multicolumn{1}{c|}{0.777}          & \multicolumn{1}{c|}{0.714}           & 0.664          & \multicolumn{1}{c|}{0.771}            & \multicolumn{1}{c|}{0.795}          & \multicolumn{1}{c|}{0.792}           & 0.727          \\ \cline{2-10} 
                                   & \emph{Similarity}                 & \multicolumn{1}{c|}{0.571}            & \multicolumn{1}{c|}{0.611}          & \multicolumn{1}{c|}{0.565}           & 0.536          & \multicolumn{1}{c|}{0.628}            & \multicolumn{1}{c|}{0.637}          & \multicolumn{1}{c|}{0.658}           & 0.59           \\ \hline
\multirow{2}{*}{\textbf{SSACL-DA}} & \emph{F1}                         & \multicolumn{1}{c|}{0.704}            & \multicolumn{1}{c|}{0.774}          & \multicolumn{1}{c|}{0.683}           & 0.643          & \multicolumn{1}{c|}{0.758}            & \multicolumn{1}{c|}{0.775}          & \multicolumn{1}{c|}{0.776}           & 0.72           \\ \cline{2-10} 
                                   & \emph{Similarity}                 & \multicolumn{1}{c|}{0.564}            & \multicolumn{1}{c|}{0.599}          & \multicolumn{1}{c|}{0.564}           & 0.63           & \multicolumn{1}{c|}{0.62}             & \multicolumn{1}{c|}{0.631}          & \multicolumn{1}{c|}{0.657}           & 0.584          \\ \hline
\multirow{2}{*}{\textbf{SSACL}}    & \emph{F1}                         & \multicolumn{1}{c|}{\textbf{0.733}}   & \multicolumn{1}{c|}{\textbf{0.802}} & \multicolumn{1}{c|}{\textbf{0.716}}  & \textbf{0.681} & \multicolumn{1}{c|}{\textbf{0.793}}   & \multicolumn{1}{c|}{\textbf{0.818}} & \multicolumn{1}{c|}{\textbf{0.811}}  & \textbf{0.751} \\ \cline{2-10} 
                                   & \emph{Similarity}                 & \multicolumn{1}{c|}{\textbf{0.585}}   & \multicolumn{1}{c|}{\textbf{0.623}} & \multicolumn{1}{c|}{\textbf{0.585}}  & \textbf{0.547} & \multicolumn{1}{c|}{\textbf{0.647}}   & \multicolumn{1}{c|}{\textbf{0.656}} & \multicolumn{1}{c|}{\textbf{0.676}}  & \textbf{0.611} \\ \hline
\end{tabular}
\end{table*}

\subsubsection{Model Configuration}
We use DeBERTa~\cite{he2020deberta} to create pre-trained word embeddings, with an embedding size of $768$ by default.
When training \methodone, the \encoder parameters are fixed, and we only optimize the parameters for the \matcher during model training.
Specifically, we use Adam~\cite{kingma2014adam} to optimize the model with a learning rate of $10^{-6}$ for $30$ training epochs.
By default, the number of positive instances $N_{pos}$ is set to $6$ while the number of negative instances $N_{neg}$ is set to $20$.
For \methodtwo, we employ the LLM, LLama3.1~\cite{touvron2023llama}, to make predictions by default.
Furthermore, the number of demonstration examples is set to $10$ by default.
This method does not require additional training time.

\subsection{Effectiveness Study}~\label{sec: effectiveness study}
\subsubsection{\Taska}~\label{sec: effectiveness on taska}
To verify the effectiveness of our proposed \methodone on the task of \taska, we include the following baselines:

\begin{itemize}[noitemsep, leftmargin=*]

\item \textbf{Unicorn~\cite{unicorn_tu2023unicorn}} -- This method unifies six data matching tasks and achieves state-of-the-art performance for the task of entity resolution. 
\item \textbf{ALITE~\cite{alite_khatiwada2022integrating}} -- ALITE matches tuples only when their values are exactly the same in their common non-missing attributes.
\item \textbf{\methodone-AE} -- A variant of \methodone, which only uses data augmentation to generate additional training examples.
\item \textbf{\methodone-DA} -- A variant of \methodone, which only uses adversarial examples to generate additional training examples. 
\end{itemize}

As described in Sec.~\ref{sec: evaluation metrics}, we use \emph{F1} to evaluate their performance on the task of \taska.
For a fair comparison, all compared methods except for ALITE are trained using supervised learning since ALITE utilizes exact matching to judge the pairwise integrability.
Specifically, we divide each dataset into training data and test data with a ratio of 70\% and 30\%. 
While the former is used to train the models, the latter is used to evaluate their performance.

\par Tables~\ref{table: balanced}, \ref{table: SE-heavy} and \ref{table: TE-heavy} present the \emph{F1} scores for all of the compared methods using balanced noise, SE-heavy noise and typographical error-heavy noise, respectively, for both dataset repositories, \emph{Real} and \emph{Join}. 
The highest scores are shown in bold. 
Observe that:
\noindent\textit{Overall effectiveness.} At first, we can see that for all sizes of datasets and different noise configurations, our proposed \methodone achieves the best performance.
Specifically, on average, a relative improvement of \methodone over Unicorn in terms of \emph{F1} is 4.6\% and 3.8\% on \emph{Real} and \emph{Join}, respectively.
Second, \methodone-AE and \methodone-DA also exhibits promising performance.
Specifically, on average, \methodone-AE outperforms Unicorn with an average relative improvement of 2.7\% and 2.4\% on \emph{Real} and \emph{Join}, respectively.
Furthermore, \methodone-DA marginally outperforms Unicorn in most cases.
Lastly, and unsurprisingly, ALITE does not perform well since any type of noise will degrade the exact matching algorithm it uses.
The above results confirm the effectiveness of our proposed \methodone, and we attribute the improvements to two factors: 1) the design of AIJNet in Sec.~\ref{sec: integration condition judgment}, which is able to distinguish the important attributes when matching two tuples; 
2) The effectiveness of the proposed data augmentation and adversarial training methods in generating useful training examples.

\noindent\textit{Effectiveness on datasets using different noise configurations.} 
By testing how these methods behave on the three different types of datasets, we find that as the ratio of typographical errors increases, the \emph{F1} score for all of the methods decrease.
For example, the average \emph{F1} of \methodone in the case of balanced noise on \emph{Real} is $0.763$, and \methodone achieves an average \emph{F1} of $0.780$ in the case of SE-heavy noise on the same dataset repository, meaning there is a relative increase of $2.22\%$, compared with the average \emph{F1} achieved in the balanced noise case. 
In contrast, the average \emph{F1} achieved by \methodone for the TE-heavy noise case on \emph{Real} is $0.733$, indicating that there is a relative decrease of $4.94\%$ compared to the average \emph{F1} achieved in the case of balanced noise on the same dataset repository, which demonstrates that, compared to semantic equivalence, typographical errors are much more difficult to correctly resolve.
One possible reason is that both Unicorn and our methods use LLMs to produce tuple representations, which is able to support semantic equivalence to some extent, but may not effectively address typographical errors.

\noindent\textit{Impact on the task \taskb.} 
Since the result of \taska has an impact on the effectiveness of \taskb, we also demonstrate how the \emph{Similarity} scores for \taskb change if we adopt different approaches to the \taska. 
Observe that in Table~\ref{table: balanced}, \ref{table: SE-heavy} and \ref{table: TE-heavy}, the \emph{Similarity} scores present a similar trend for \emph{F1} scores, which demonstrates that an accurate prediction in \taska will also improve \taskb. 

\noindent\textit{Impact of Data Characteristics}
We observe that both numerical attributes and missing attributes negatively impact the performance of \methodone because numerical data are more challenging to capture semantically than textual attributes, and missing attributes provide no useful information for the decision. Nevertheless, even when both $\alpha_1$ and $\alpha_2$ are high, our method still achieves strong F1 scores and Similarity, demonstrating its robustness under high ratios of numerical and missing attributes.

\begin{table}[h]
\centering
\footnotesize
\caption{The impact of the ratio of numerical attributes on \emph{F1} and \emph{Similarity}}\label{tab: numerical attributes}
\begin{tabular}{|c|c|c|c|c|}
\hline
\textbf{$\alpha_1$} & \textbf{5-10\%} & \textbf{10-15\%} & \textbf{15-20\%} & \textbf{20-25\%} \\ \hline
\textit{F1}         & 0.890             & 0.836              & 0.774              & 0.745              \\ \hline
\textit{Similarity} & 0.723             & 0.675              & 0.649              & 0.628              \\ \hline
\end{tabular}
\end{table}

\begin{table}[h]
\centering
\footnotesize
\caption{The impact of the ratio of missing values on \emph{F1} and \emph{Similarity}}\label{tab: missing values}
\begin{tabular}{|c|c|c|c|c|}
\hline
\textbf{$\alpha_2$} & \textbf{0-15\%} & \textbf{15-30\%} & \textbf{30-45\%} & \textbf{$\geq$45\%} \\ \hline
\textit{F1}         & 0.839            & 0.818            & 0.783            & 0.765         \\ \hline
\textit{Similarity} & 0.704            & 0.679            & 0.658            & 0.647         \\ \hline
\end{tabular}
\end{table}

\subsubsection{\Taskb}~\label{sec: effectiveness on taskb} 
We conduct an empirical study to explore which of the methods proposed in Sec.~\ref{sec: integrable set identification} are effective for the task \taskb. 
To achieve this, we compare the following methods: the Bron-Kerbosch algorithm (BK)~\cite{regneri2007bronalg}, Louvain~\cite{louvain}, the Newman-Girvan algorithm~\cite{newman2004NGalg} (NG), Informap~\cite{rosvall2008infomap}, Spectral Clustering (SC)~\cite{newman2013spectral}, and a Graph Neural Network(GNN)~\cite{bruna2017community}, which were briefly introduced in Sec.~\ref{def: integrable set discovery}.

Table~\ref{table: effectiveness on taskb} demonstrates that the effectiveness of the compared methods for the two datasets, in terms of \emph{Similarity}. 
Observe that:
\begin{itemize}[noitemsep, leftmargin=*]
\item We can see that for all methods, a GNN achieves the best performance in terms of \emph{Similarity}, outperforming all other methods by a large margin.
Specifically, when compared against the second best method, spectral clustering, the relative improvement for a GNN on \emph{Real} and \emph{Join} are 6.6\% and 7.5\% on average, respectively.
We attribute this improvement to the fact that a GNN learns more accurate node representations by employing a non-linear aggregation operation, which effectively aggregates information from neighboring nodes.
\item All community detection methods outperform the maximal clique identification method, Bron-Kerbosch algorithm (BK).
Even Louvain, which achieves the lowest \emph{Similarity} for all of the community detection algorithm, outperforms the Bron-Kerbosch algorithm with a relative improvement of 21.7\% and 27.6\% on \emph{Real} and \emph{Join}, respectively.
This verifies that maximal clique identification methods are not robust for the task \taskb, since it strictly requires that every tuple pair in the integrable set should be decided to be integrable by the \taska method, which is difficult in practice.
\end{itemize}

\begin{table*}[pt]
\centering
\caption{Effectiveness of \taskb approaches for the \emph{Real} and \emph{Join} datasets}~\label{table: effectiveness on taskb}
\begin{tabular}{|c|cccc|cccc|}
\hline
\multirow{2}{*}{\textbf{Methods}} & \multicolumn{4}{c|}{\textbf{Real}}                                                                                                  & \multicolumn{4}{c|}{\textbf{Join}}                                                                                                  \\ \cline{2-9} 
                                  & \multicolumn{1}{c|}{\textbf{Overall}} & \multicolumn{1}{c|}{\textbf{Small}} & \multicolumn{1}{c|}{\textbf{Medium}} & \textbf{Large} & \multicolumn{1}{c|}{\textbf{Overall}} & \multicolumn{1}{c|}{\textbf{Small}} & \multicolumn{1}{c|}{\textbf{Medium}} & \textbf{Large} \\ \hline
\textbf{BK}                       & \multicolumn{1}{c|}{0.386}            & \multicolumn{1}{c|}{0.393}          & \multicolumn{1}{c|}{0.372}           & 0.364          & \multicolumn{1}{c|}{0.399}            & \multicolumn{1}{c|}{0.384}          & \multicolumn{1}{c|}{0.426}           & 0.371          \\ \hline
\textbf{Louvain}                  & \multicolumn{1}{c|}{0.465}            & \multicolumn{1}{c|}{0.486}          & \multicolumn{1}{c|}{0.474}           & 0.423          & \multicolumn{1}{c|}{0.507}            & \multicolumn{1}{c|}{0.53}           & \multicolumn{1}{c|}{0.517}           & 0.461          \\ \hline
\textbf{GN}                       & \multicolumn{1}{c|}{0.479}            & \multicolumn{1}{c|}{0.512}          & \multicolumn{1}{c|}{0.498}           & 0.447          & \multicolumn{1}{c|}{0.522}            & \multicolumn{1}{c|}{0.558}          & \multicolumn{1}{c|}{0.543}           & 0.487          \\ \hline
\textbf{SC}                       & \multicolumn{1}{c|}{0.572}            & \multicolumn{1}{c|}{0.586}          & \multicolumn{1}{c|}{0.568}           & 0.564          & \multicolumn{1}{c|}{0.623}            & \multicolumn{1}{c|}{0.639}          & \multicolumn{1}{c|}{0.619}           & 0.615          \\ \hline
\textbf{Infomap}                  & \multicolumn{1}{c|}{0.562}            & \multicolumn{1}{c|}{0.573}          & \multicolumn{1}{c|}{0.547}           & 0.513          & \multicolumn{1}{c|}{0.613}            & \multicolumn{1}{c|}{0.625}          & \multicolumn{1}{c|}{0.596}           & 0.559          \\ \hline
\textbf{GNN}                      & \multicolumn{1}{c|}{\textbf{0.610}}   & \multicolumn{1}{c|}{\textbf{0.648}} & \multicolumn{1}{c|}{\textbf{0.608}}  & \textbf{0.574} & \multicolumn{1}{c|}{\textbf{0.67}}    & \multicolumn{1}{c|}{\textbf{0.697}} & \multicolumn{1}{c|}{\textbf{0.683}}  & \textbf{0.632} \\ \hline
\end{tabular}
\end{table*}

\begin{table*}[pt]
\centering
\caption{Effectiveness of \taskc approaches on the \emph{Real} and \emph{Join} datasets}~\label{sec: effectiveness on taskc}
\begin{tabular}{|c|cccc|cccc|}
\hline
\multirow{2}{*}{\textbf{Methods}} & \multicolumn{4}{c|}{\textbf{Real}}                                                                                                  & \multicolumn{4}{c|}{\textbf{Join}}                                                                                                  \\ \cline{2-9} 
                                  & \multicolumn{1}{c|}{\textbf{Overall}} & \multicolumn{1}{c|}{\textbf{Small}} & \multicolumn{1}{c|}{\textbf{Medium}} & \textbf{Large} & \multicolumn{1}{c|}{\textbf{Overall}} & \multicolumn{1}{c|}{\textbf{Small}} & \multicolumn{1}{c|}{\textbf{Medium}} & \textbf{Large} \\ \hline
\textbf{Random}                   & \multicolumn{1}{c|}{0.266}            & \multicolumn{1}{c|}{0.287}          & \multicolumn{1}{c|}{0.251}           & 0.254          & \multicolumn{1}{c|}{0.298}            & \multicolumn{1}{c|}{0.285}          & \multicolumn{1}{c|}{0.325}           & 0.279          \\ \hline
\textbf{Major}                    & \multicolumn{1}{c|}{0.317}            & \multicolumn{1}{c|}{0.331}          & \multicolumn{1}{c|}{0.305}           & 0.318          & \multicolumn{1}{c|}{0.357}            & \multicolumn{1}{c|}{0.394}          & \multicolumn{1}{c|}{0.332}           & 0.365          \\ \hline
\textbf{SlimFast}                 & \multicolumn{1}{c|}{0.626}            & \multicolumn{1}{c|}{0.658}          & \multicolumn{1}{c|}{0.614}           & 0.603          & \multicolumn{1}{c|}{0.641}            & \multicolumn{1}{c|}{0.625}          & \multicolumn{1}{c|}{0.660}           & 0.649          \\ \hline
\textbf{ICLCR}                    & \multicolumn{1}{c|}{\textbf{0.724}}   & \multicolumn{1}{c|}{\textbf{0.735}} & \multicolumn{1}{c|}{\textbf{0.726}}  & \textbf{0.709} & \multicolumn{1}{c|}{\textbf{0.737}}   & \multicolumn{1}{c|}{\textbf{0.734}} & \multicolumn{1}{c|}{\textbf{0.759}}  & \textbf{0.728} \\ \hline
\end{tabular}
\end{table*}

\subsubsection{\Taskc}\label{sec: effectivess on taskc}
For task~\taskc, we compare our proposed \methodtwo against the following methods: 
\begin{itemize}[noitemsep, leftmargin=*]
\item \textbf{\rand} -- A simple baseline that resolves conflicts by uniformly selecting a value from the candidate set at random.
\item \textbf{\major} -- A baseline that resolves conflicts by selecting the most frequent value from the candidate set using kNN.
\item \textbf{\slim}~\cite{sigmod_df_rekatsinas2017slimfast} -- A state-of-the-art truth discovery model for conflict resolution for single fact scenarios using weighted kNN. 
\end{itemize} 
To ensure a fair comparison,  we balance the labeled data available for \slim and \methodtwo. 
In \slim, the labeled data is used to train the model, whereas in \methodtwo, the labeled data is used as demonstration examples. 
Specifically, for \methodtwo, the maximum number of demonstration examples $N_{max}$ is achieved when the prompt input exactly reaches the maximum input size of the LLM.
Thus, for each dataset, we first identify $N_{max}$, which is also used as the training data for \slim. 
\par Table~\ref{sec: effectiveness on taskc} demonstrates the effectiveness for all of the methods using the \emph{Real} and \emph{Join} datasets. 
Observe that: our proposed \methodtwo achieves the highest \emph{Accuracy} in every case. 
Specifically, the overall \emph{Accuracy} of \methodtwo for \emph{Real} and \emph{Join} is 0.724 and 0.737, respectively, surpassing \slim by margins of 14.3\% and 22.6\%, respectively.
Furthermore, both \rand and \major perform poorly, as they rely on heuristics for the candidate value selection. 
This highlights the effectiveness of the \methodtwo method, which can be attributed to the fact that \methodtwo fully leverages knowledge embedded in an LLM, and gains insights from the selected demonstration examples to make the prediction.

\subsection{End-to-end Effectiveness on Downstream Tasks}\label{sec: end-to-end}
In this experiment, we evaluate whether the integrated table generated by our method can significantly improve performance compared to the base table. To achieve this, we selected three datasets—R4, R9, and J11—from the \emph{Real} and \emph{Join} repositories, each containing a base table suitable for a classification task.
First, for each dataset, we randomly split the data from the base table into training, validation, and test sets in a ratio of 7:2:1. We then trained a three-layer multi-layer perceptron (MLP) on the training set and evaluated its performance on the test set to establish baseline performance, measured by \emph{Accuracy}.
Next, we generated an augmented table for each dataset using two integration pipelines:
\begin{itemize}[leftmargin=*,noitemsep] \item \emph{BBL:} A combination of the \underline{b}est \underline{b}ase\underline{l}ine methods from previous experiments, including Unicorn, GNN, and \slim, to generate the augmented table.
\item \emph{Our:} A combination of our proposed methods, including \methodone, GNN, and \methodtwo, to generate the augmented table.
\end{itemize}
For each augmented table, we applied the same data splitting, training, and evaluation settings as the base table. Table~\ref{tab: downstream tasks} presents the \emph{Accuracy} achieved by the same model on the different tables. As shown, using augmented tables for the classification task generally leads to an improvement in model performance. Moreover, the model trained on the augmented table generated by our integration pipeline consistently outperformed the model trained on the augmented table generated by the baseline pipeline. Specifically, the relative improvements in \emph{Accuracy} on the R4, R9, and J11 datasets are 0.8\%, 2.21\%, and 2.71\%, respectively. These results demonstrate that our proposed integration pipeline provides more accurate and comprehensive integration results, effectively enhancing model performance.

\begin{table}[h]
\centering
\caption{The effectiveness of different table integration pipelines for classification tasks}\label{tab: downstream tasks}
\begin{tabular}{|c|ccc|}
\hline
\multicolumn{1}{|l|}{\multirow{2}{*}{}} & \multicolumn{3}{c|}{\textbf{Accuracy}}                                                                          \\ \cline{2-4} 
\multicolumn{1}{|l|}{}                  & \multicolumn{1}{c|}{\textbf{R4}}    & \multicolumn{1}{c|}{\textbf{R9}}    & \textbf{J11}                        \\ \hline
\textbf{Base}                           & \multicolumn{1}{c|}{0.735}          & \multicolumn{1}{c|}{0.672}          & 0.782                               \\ \hline
\textbf{BBL}                            & \multicolumn{1}{c|}{0.746}          & \multicolumn{1}{c|}{0.679}          & 0.774                               \\ \hline
\textbf{Our}                            & \multicolumn{1}{l|}{\textbf{0.752}} & \multicolumn{1}{l|}{\textbf{0.694}} & \multicolumn{1}{l|}{\textbf{0.795}} \\ \hline
\end{tabular}
\end{table}

\subsection{Label Efficiency Study}~\label{sec: label-efficiency study}
\subsubsection{\Taska}
As introduced in Sec.~\ref{sec: integration condition judgment}, one significant advantage of \methodone is its ability to be efficiently trained using limited labeled data, or in a fully self-supervised manner.
Therefore, we also conduct an experiment to verify the effectiveness of \methodone using minimal labeled data.
We compare the performance of \methodone in both supervised and self-supervised settings.
Furthermore, since \methodone's ability to perform self-supervised learning primarily stems from the proposed data augmentation and adversarial training techniques, we also explore different approaches to self-supervised learning.
Specifically, we compare the following training for \methodone:
\begin{itemize}[noitemsep, leftmargin=*]
\item \textit{SL} -- it trains \methodone in a supervised manner, as introduced in Sec.~\ref{sec: effectiveness on taska}.
\item \textit{SSL} -- it trains \methodone in a self-supervised manner, using both data augmentation and adversarial training to generate the training examples.
\item \textit{SSL-AE} -- its trains \methodone in a self-supervised manner, using only data augmentation to generate training examples.
\end{itemize}

Note that we can not only use adversarial training to generate training examples for self-supervised training, the adversarial training examples are identified using an existing positive training pair.

\begin{table*}[th]
\centering
\caption{Effectiveness of \methodone on \taska using different training strategies}~\label{table: training strategies}
\begin{tabular}{|c|c|cccc|cccc|}
\hline
\multirow{2}{*}{\textbf{Methods}} & \multirow{2}{*}{\textbf{Metrics}} & \multicolumn{4}{c|}{\textbf{Real}}                                                                                                  & \multicolumn{4}{c|}{\textbf{Join}}                                                                                                  \\ \cline{3-10} 
                                  &                                   & \multicolumn{1}{c|}{\textbf{Overall}} & \multicolumn{1}{c|}{\textbf{Small}} & \multicolumn{1}{c|}{\textbf{Medium}} & \textbf{Large} & \multicolumn{1}{c|}{\textbf{Overall}} & \multicolumn{1}{c|}{\textbf{Small}} & \multicolumn{1}{c|}{\textbf{Medium}} & \textbf{Large} \\ \hline
\multirow{2}{*}{\textbf{SL}}      & \emph{F1}                         & \multicolumn{1}{c|}{0.718}            & \multicolumn{1}{c|}{0.774}          & \multicolumn{1}{c|}{0.696}           & 0.674          & \multicolumn{1}{c|}{0.785}            & \multicolumn{1}{c|}{0.813}          & \multicolumn{1}{c|}{0.784}           & 0.741          \\ \cline{2-10} 
                                  & \emph{Similarity}                 & \multicolumn{1}{c|}{0.570}            & \multicolumn{1}{c|}{0.611}          & \multicolumn{1}{c|}{0.566}           & 0.535          & \multicolumn{1}{c|}{0.627}            & \multicolumn{1}{c|}{0.662}          & \multicolumn{1}{c|}{0.643}           & 0.594          \\ \hline
\multirow{2}{*}{\textbf{SSL-AE}}  & \emph{F1}                         & \multicolumn{1}{c|}{0.709}            & \multicolumn{1}{c|}{0.767}          & \multicolumn{1}{c|}{0.688}           & 0.664          & \multicolumn{1}{c|}{0.764}            & \multicolumn{1}{c|}{0.791}          & \multicolumn{1}{c|}{0.775}           & 0.731          \\ \cline{2-10} 
                                  & \emph{Similarity}                 & \multicolumn{1}{c|}{0.563}            & \multicolumn{1}{c|}{0.597}          & \multicolumn{1}{c|}{0.561}           & 0.529          & \multicolumn{1}{c|}{0.622}            & \multicolumn{1}{c|}{0.647}          & \multicolumn{1}{c|}{0.631}           & 0.588          \\ \hline
\multirow{2}{*}{\textbf{SSL}}     & \emph{F1}                         & \multicolumn{1}{c|}{\textbf{0.763}}   & \multicolumn{1}{c|}{\textbf{0.831}} & \multicolumn{1}{c|}{\textbf{0.744}}  & \textbf{0.715} & \multicolumn{1}{c|}{\textbf{0.827}}   & \multicolumn{1}{c|}{\textbf{0.856}} & \multicolumn{1}{c|}{\textbf{0.84}}   & \textbf{0.787} \\ \cline{2-10} 
                                  & \emph{Similarity}                 & \multicolumn{1}{c|}{\textbf{0.610}}   & \multicolumn{1}{c|}{\textbf{0.648}} & \multicolumn{1}{c|}{\textbf{0.608}}  & \textbf{0.574} & \multicolumn{1}{c|}{\textbf{0.67}}    & \multicolumn{1}{c|}{\textbf{0.697}} & \multicolumn{1}{c|}{\textbf{0.682}}  & \textbf{0.632} \\ \hline
\end{tabular}
\end{table*}

Table~\ref{table: training strategies} shows the experimental results.
Unsurprisingly, SL achieves the best performance for all cases since it uses enough training examples to train \methodone.
However, SSL also achieves comparable performance.
Specifically, on average, the relative performance of SSL over SL is 0.932\% and 0.947\% on \emph{Real} and \emph{Join}, respectively. 
This demonstrates the effectiveness of our proposed method when automatically generating labeled data, which enables our model to achieve comparable performance to the model trained using labeled data.
Furthermore, SSL also performs slightly better than SSL-AE, which confirms the value of introducing adversarial examples to enhance model training.
Overall, the experimental results verify the ability of \methodone to train using minimal labeled data.

\subsubsection{\Taskc}\label{sec: label efficiency on taskc}
\begin{figure}[h]
    \centering
    \includegraphics[width=\linewidth]{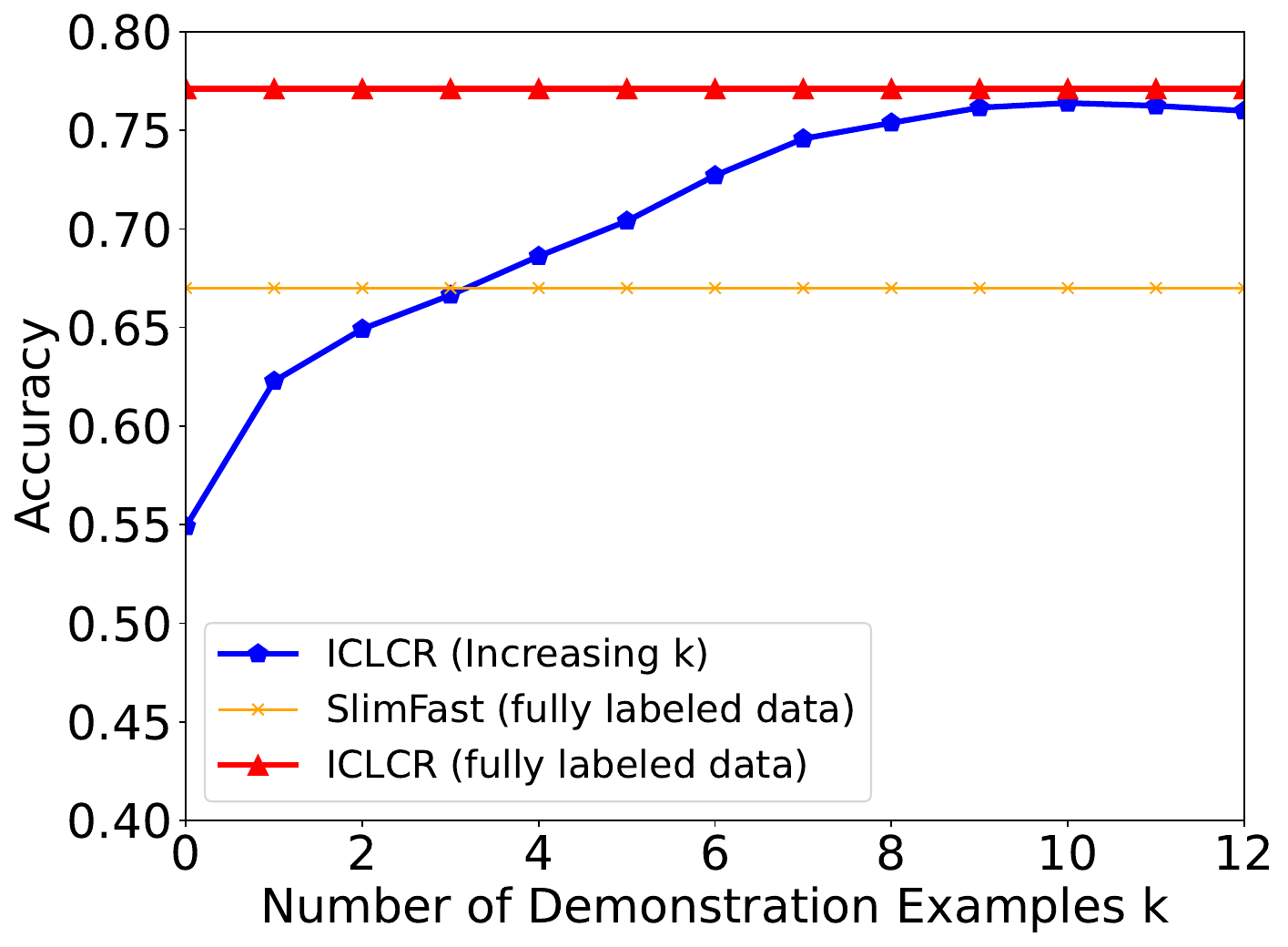}
    \caption{Impact of the number of demonstration examples $k$ on the \emph{Accuracy} of \methodtwo}
    \label{fig: exp on k}
\end{figure}

As introduced in Sec.~\ref{sec: multi-tuple integration}, the main advantage of \methodtwo is that the model can perform well even with limited labeled data. 
To verify this, we study if \methodtwo can maintain competitive performance even when using a constrained number of demonstration examples in this experiment.
Specifically, we test the performance of \methodtwo on the J28 dataset, progressively increasing the number of demonstration examples $k$ from 0 to 12.
The results are presented in Fig.~\ref{fig: exp on k}.
We also illustrate \emph{Accuracy} of \slim and \methodtwo using a sufficient amount labeled data for supervised training (as introduced in Sec.~\ref{sec: effectiveness on taskc}) in the figure as a reference.

\par When $k=0$, \methodtwo does not rely on any labeled data to make predictions. 
However, it still achieves an accuracy of 0.550. This demonstrates that even when there are no labeled data, \methodtwo can still leverage the knowledge of a large training corpus in an LLM to achieve relatively high performance.
As $k$ increases, a consistent improvement in model performance is observed, as the increased number of demonstration examples guide \methodtwo decisions, enhancing the prediction accuracy.
Specifically, when $k=3$, \methodtwo achieves performance comparable to that of \slim.

When $k=12$, \methodtwo achieves a 98.2\% accuracy, compared to \methodtwo using a complete set of labeled data (Sec.~\ref{sec: effectiveness on taskc}).
This improvement trend illustrates that by increasing the number of demonstration examples, \methodtwo can acquire new knowledge from examples and improve the performance for the task~\taskc.
This further verifies the effectiveness of \methodtwo in a scenario where the availability of labeled data is limited.

\begin{table*}[tp]
\centering
\caption{Impact of using different pre-trained LLMs for integrable set discovery}~\label{table: llms1}
\begin{tabular}{|c|c|c|c|c|c|c|}
\hline
                               &                   & \textbf{BERT} & \textbf{RoBERTa} & \textbf{DistilBERT} & \textbf{XLNet} & \textbf{DeBERTa} \\ \hline
\multirow{2}{*}{\textbf{Real}} & \emph{F1}         & 0.758         & 0.760            & 0.742               & 0.751          & \textbf{0.763}   \\ \cline{2-7} 
                               & \emph{Similarity} & 0.607         & 0.609            & 0.582               & 0.598          & \textbf{0.610}   \\ \hline
\multirow{2}{*}{\textbf{Join}} & \emph{F1}         & 0.824         & 0.823            & 0.801               & 0.817          & \textbf{0.827}   \\ \cline{2-7} 
                               & \emph{Similarity} & 0.665         & 0.668            & 0.637               & 0.654          & \textbf{0.670}   \\ \hline
\end{tabular}
\end{table*}

\subsection{Impact of the Choices of PLMs and LLMs}\label{sec: LLM Choice Study}
\subsubsection{\Taska}~\label{sec: LLM choice on taska}
For the task of \taska, we primarily use PLMs to initialize the word embeddings for tokens in a tuple. 
In this experiment, we investigate how different choices of PLMs impact the method performance for both tasks of \taska. 
Specifically, we include five widely used PLMs in the comparison, namely BERT~\cite{devlin2019bert}, RoBERTa~\cite{liu2019roberta}, DistilBERT~\cite{sanh2019distilbert}, XLNet~\cite{yang2019xlnet}, and DeBERTa~\cite{he2020deberta}, and the comparison result is shown in Table~\ref{table: llms1}. 
Among all the pre-trained PLMs tested, DeBERTa achieves the best performance for the two dataset repositories, \emph{Real} and \emph{Join}, in terms of \emph{F1} and \emph{Similarity}.
One possible reason for this is that DeBERTa employs a new positional encoding strategy that captures relative position information of the tokens in a sequence. 
Additionally, BERT and RoBERTa also perform well on the two tasks, exhibiting similar performance. 
Lastly, DistilBERT achieves the worst \emph{F1} and \emph{Similarity} scores, possibly because it has a much smaller model, where the information included is distilled from larger models.

\subsubsection{\Taskc}\label{sec: LLM choice on taskc}
In \methodtwo, we primarily employ large language models (LLMs) to select the candidate value that has the highest probability for the given prompt input.
In this section, we conduct experiments to investigate how the different choice of LLMs has an impact on the performance on \taskb. 
To achieve this, we select three open-source and up-to-date LLMs, Qwen 2 (7B)~\cite{bai2023qwen2}, Mistral~\cite{jiang2023mistral}, and LLama 3.1 (8B)~\cite{dubey2024llama3} since they are widely used in the field of in-context learning.
Furthermore, we evaluate their performance on \emph{Real} and \emph{Join} benchmarks.
The results are shown in Table~\ref{table: llms2}. 
Notably, LLama 3.1 achieves the highest average accuracy across the two benchmarks, while Mistral and Qwen 2 also deliver competitive results. Looking ahead, equipping \methodtwo with more advanced and powerful LLMs is expected to further enhance its performance. 

\begin{table}[h]
\centering

\caption{Impact of using different LLMs for the task of \taskc}~\label{table: llms2}

\begin{tabular}{|cc|c|c|c|}
\hline
\multicolumn{2}{|c|}{}                                                  & \textbf{Qwen 2} & \textbf{Mistral} & \textbf{LLama 3.1} \\ \hline
\multicolumn{1}{|c|}{\multirow{4}{*}{\textbf{Real}}} & \textbf{Overall} & 0.710          & 0.715            & \textbf{0.724}    \\ \cline{2-5} 
\multicolumn{1}{|c|}{}                               & \textbf{Small}   & 0.713          & 0.714            & \textbf{0.735}    \\ \cline{2-5} 
\multicolumn{1}{|c|}{}                               & \textbf{Medium}  & 0.709          & \textbf{0.727}   & 0.726             \\ \cline{2-5} 
\multicolumn{1}{|c|}{}                               & \textbf{Large}   & \textbf{0.716} & 0.708            & 0.709             \\ \hline
\multicolumn{1}{|c|}{\multirow{4}{*}{\textbf{Join}}} & \textbf{Overall} & 0.723          & 0.712            & \textbf{0.737}    \\ \cline{2-5} 
\multicolumn{1}{|c|}{}                               & \textbf{Small}   & 0.717          & 0.708            & \textbf{0.734}    \\ \cline{2-5} 
\multicolumn{1}{|c|}{}                               & \textbf{Medium}  & 0.728          & 0.724            & \textbf{0.759}    \\ \cline{2-5} 
\multicolumn{1}{|c|}{}                               & \textbf{Large}   & 0.724          & 0.707            & \textbf{0.728}    \\ \hline
\end{tabular}
\end{table}

\subsection{Hyper-parameters}\label{sec: hyper-parameter study}
\subsubsection{Impact of The Number of Positive and Negative Instances}\label{sec: taska hyperparameter1}
\begin{figure}[h]
    \centering

    \subfloat[Number of Positive Instances]{%
        \includegraphics[width=0.24\textwidth]{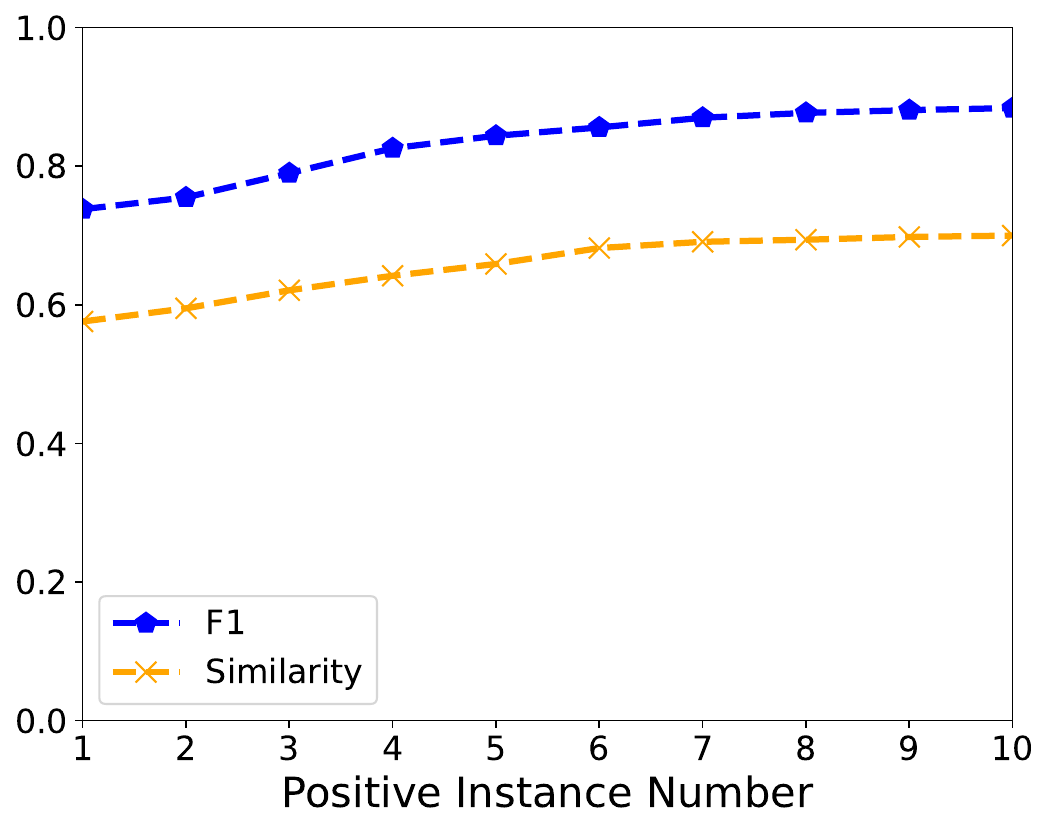}%
        \label{fig: pos}%
    }
    \hfill
    \subfloat[Number of Negative Instances]{%
        \includegraphics[width=0.24\textwidth]{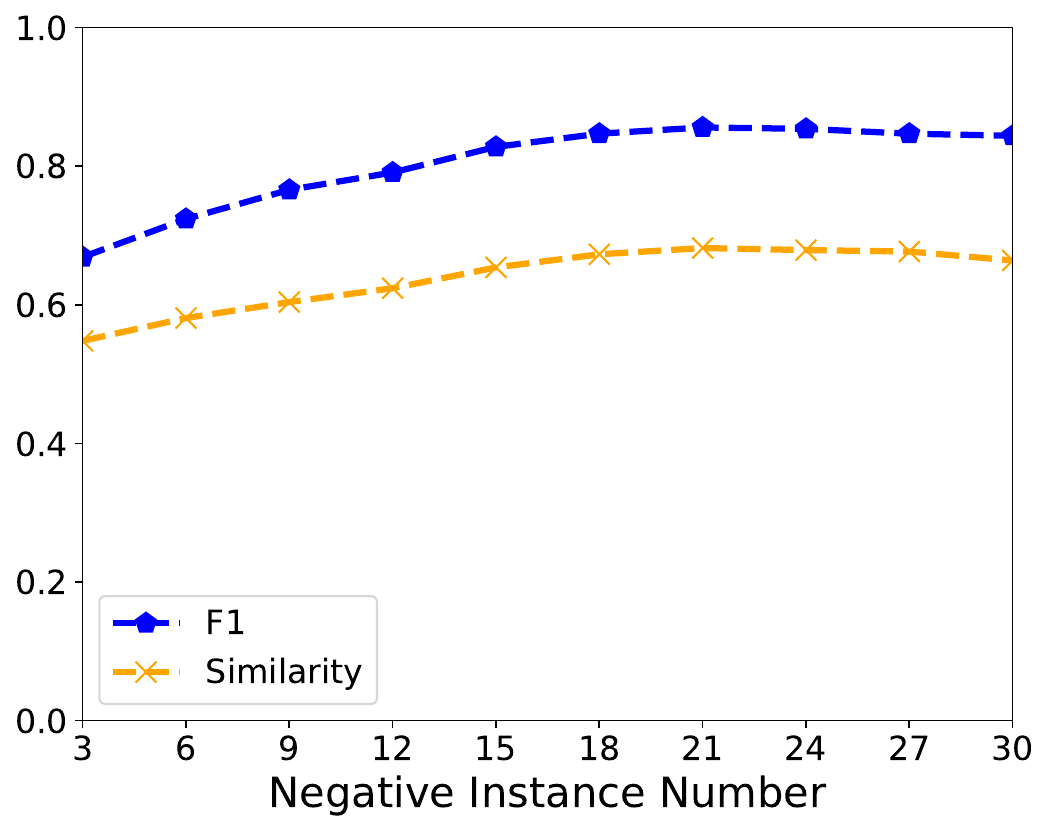}%
        \label{fig: neg}%
    }
    \captionsetup{skip=3pt}
    \caption{The Impact of Positive and Negative Instances on \emph{F1} and \emph{Similarity} }

    \label{fig: pos and neg}
\end{figure}

In this experiment, we investigate the impact of two key hyper-parameters -- the number of positive instances $N_{pos}$ and the number of negative instances $N_{neg}$, on the model performance of \methodone.
We use $R11$, the largest dataset from \emph{Real}, to evaluate the model performance as this hyper-parameter is varied.
The experimental results are shown in Fig.~\ref{fig: pos and neg}.
\par Observe that:
As $N_{pos}$ increases, the performance of \methodone consistently improves.
This is because the proposed data augmentation methods generate more diverse types of positive instances, enabling the model to capture a range of semantic equivalence and typographical errors.
However, when $N_{pos}$ exceeds a certain threshold, namely 6, the model performance improvements are marginal.
Since increasing $N_{pos}$ also requires additional model training time, we recommend setting $N_{pos}$ to 6.

\par In contrast to $N_{pos}$, whose improvements are easy to see, increasing $N_{neg}$ yields better models only within a range from 3 to 21.
Within this range, a greater $N_{neg}$ improves the model.
However, as $N_{neg}$ exceeds 21, the model performance suddenly begins to deteriorate. 
This phenomenon can be attributed to the likelihood of including positive instances as negative training data when $N_{neg}$ becomes exceptionally large, thereby reducing model performance. Therefore, we recommend to set $N_{neg}$ to 20. 

\subsection{Ablation Study}~\label{sec: ablation study}
\subsubsection{Effectiveness of Demonstration Example Compression Strategy}\label{sec: exp on dec}
In this experiment, we investigate the impact of the proposed demonstration example compression strategy on the performance of \methodtwo.
In particular, we compare the standard \methodtwo against \methodtwo-DEC, a version of \methodtwo that does not include the demonstration example compression strategy. 
As shown in Table~\ref{table: exp on dec}, \methodtwo outperforms \methodtwo-DEC, with a relative improvement of 6.57\% and 7.51\% on \emph{Real} and \emph{Join}, respectively, which demonstrates the effectiveness of the proposed demonstration example compression strategy.

\begin{table*}[tp]
\centering
\caption{Impact of the demonstration compression strategy on \taskc}\label{table: exp on dec}
\begin{tabular}{|c|cccc|cccc|}
\hline
\multirow{2}{*}{\textbf{Real}} & \multicolumn{4}{c|}{\textbf{Join}}                                                                                                  & \multicolumn{4}{c|}{\textbf{Join}}                                                                                                  \\ \cline{2-9} 
                               & \multicolumn{1}{c|}{\textbf{Overall}} & \multicolumn{1}{c|}{\textbf{Small}} & \multicolumn{1}{c|}{\textbf{Medium}} & \textbf{Large} & \multicolumn{1}{c|}{\textbf{Overall}} & \multicolumn{1}{c|}{\textbf{Small}} & \multicolumn{1}{c|}{\textbf{Medium}} & \textbf{Large} \\ \hline
\textbf{ICLCR-DEC}             & \multicolumn{1}{c|}{0.689}            & \multicolumn{1}{c|}{0.670}          & \multicolumn{1}{c|}{0.677}           & 0.669          & \multicolumn{1}{c|}{0.692}            & \multicolumn{1}{c|}{0.698}          & \multicolumn{1}{c|}{0.701}           & 0.686          \\ \hline
\textbf{ICLCR}                 & \multicolumn{1}{c|}{\textbf{0.724}}   & \multicolumn{1}{c|}{\textbf{0.735}} & \multicolumn{1}{c|}{\textbf{0.726}}  & \textbf{0.709} & \multicolumn{1}{c|}{\textbf{0.737}}   & \multicolumn{1}{c|}{\textbf{0.734}} & \multicolumn{1}{c|}{\textbf{0.759}}  & \textbf{0.728} \\ \hline
\end{tabular}
\end{table*}

\subsubsection{Impact of Demonstration Example Selection Strategies}~\label{sec: exp on des}
We also compare three different demonstration example selection strategies, Random, $k$-NN, and weighted $k$-NN in \methodtwo to investigate the impact on model performance.
As shown in Table~\ref{table: exp on selection strategy}, we can first observe that random selection of demonstration examples does not fully exploit the performance of \methodtwo.
This is shown by \methodtwo performing worse with a random selection strategy, compared to the two $k$-NN strategies, which demonstrates the importance of selecting relevant demonstration examples to instruct the LLM for prediction in \methodtwo. 
Furthermore, \methodtwo with the weighted $k$-NN selection strategy achieves the highest \emph{Accuracy} in every case.
On average, \methodtwo with the weighted $k$-NN selection strategy outperforms \methodtwo with $k$-NN selection strategy by 1.1\% and 1.9\% on \emph{Real} and \emph{Join}, respectively.
This is because the weighted $k$-NN selection strategy considers the importance of different attributes for a target attribute.

\begin{table*}[tp]
\centering
\caption{The impact of the demonstration selection strategies on \taskc}~\label{table: exp on selection strategy}
\begin{tabular}{|c|cccc|cccc|}
\hline
\multirow{2}{*}{\textbf{Methods}} & \multicolumn{4}{c|}{\textbf{Real}}                                                                                                  & \multicolumn{4}{c|}{\textbf{Join}}                                                                                                  \\ \cline{2-9} 
                                  & \multicolumn{1}{c|}{\textbf{Overall}} & \multicolumn{1}{c|}{\textbf{Small}} & \multicolumn{1}{c|}{\textbf{Medium}} & \textbf{Large} & \multicolumn{1}{c|}{\textbf{Overall}} & \multicolumn{1}{c|}{\textbf{Small}} & \multicolumn{1}{c|}{\textbf{Medium}} & \textbf{Large} \\ \hline
\textbf{Random}                   & \multicolumn{1}{c|}{0.687}            & \multicolumn{1}{c|}{0.694}          & \multicolumn{1}{c|}{0.708}           & 0.692          & \multicolumn{1}{c|}{0.723}            & \multicolumn{1}{c|}{0.718}          & \multicolumn{1}{c|}{0.706}           & 0.714          \\ \hline
\textbf{k-NN}                     & \multicolumn{1}{c|}{0.709}            & \multicolumn{1}{c|}{0.704}          & \multicolumn{1}{c|}{0.716}           & 0.705          & \multicolumn{1}{c|}{0.730}            & \multicolumn{1}{c|}{0.726}          & \multicolumn{1}{c|}{0.714}           & 0.721          \\ \hline
\textbf{Weighted k-NN}            & \multicolumn{1}{c|}{\textbf{0.724}}   & \multicolumn{1}{c|}{\textbf{0.735}} & \multicolumn{1}{c|}{\textbf{0.726}}  & \textbf{0.709} & \multicolumn{1}{c|}{\textbf{0.737}}   & \multicolumn{1}{c|}{\textbf{0.734}} & \multicolumn{1}{c|}{\textbf{0.759}}  & \textbf{0.728} \\ \hline
\end{tabular}
\end{table*}

\subsection{Efficiency Study}
In this section, we mainly employ the largest dataset, namely J28,  among \emph{Real} and \emph{Join} dataset repository to conduct the efficiency study. Specifically, J28 consists of more than 99,000 rows and 26 attributes.

\subsubsection{\Taska}\label{sec: efficiency study on taska}
Overall, it takes \methodone and Unicorn 103ms and 126ms to judge the integrability for each tuple pair on average.
The average inference time of the two methods are similar, since both of them rely on a pre-trained language model (PLM) to construct the tuple representations and make prediction. Furthermore, the training time for \methodone is 10.7 hours, while the total inference time for SSACL to complete the \taska task on J28 is 5.9 hours, following the application of the LSH-based blocking mechanism~\cite{deeper_ebraheem2018distributed}.  Furthermore, the memory footprint for \methodone is 789MB of RAM and 7.6GB of VRAM.

\subsubsection{\Taskb}\label{sec: efficiency on taskb}
 In this experiment, we conduct an efficiency study to compare the time costs of various methods on the \taskb. 
Table~\ref{sec: efficieny on taskb} shows the running time cost for these methods using the J28 dataset. 
Observe that the GNN incurs the highest time costs due to the computationally intensive aggregation operations used.
Considering that GNN delivers optimal performance for the \taskb and the time required for \taskb is significantly less than \taska, GNNs are still the preferred choice for \taskb.

\begin{table}[h]
\centering
\caption{The efficiency of different \taskb methods on R11 dataset}~\label{sec: efficieny on taskb}
\begin{tabular}{|c|c|}
\hline
\multirow{2}{*}{\textbf{Methods}} & \multirow{2}{*}{\textbf{Time (Hours)}} \\
                                  &                                             \\ \hline
\textbf{BK}                       & 3.41                                        \\ \hline
\textbf{Louvain}                  & 2.62                                        \\ \hline
\textbf{GN}                       & 3.37                                        \\ \hline
\textbf{SC}                       & 3.64                                        \\ \hline
\textbf{Infomap}                  & 2.35                                        \\ \hline
\textbf{GNN}                      & 3.91                                        \\ \hline
\end{tabular}
\end{table}

\subsubsection{\Taskc} 
On average, it takes \methodtwo and \slim 35ms and 417ms to make predictions for each integrable set in the dataset $J28$.
Our approach is more costly for two reasons: 
(1) \methodtwo relies on an LLM to make the predictions, which has higher model complexity than \slim; 
(2) Our model is based on in-context learning, which requires the model to input each text test case as well as any demonstration examples.
Thus, a larger input size has higher running time costs.
However, considering that the relative performance improvement for \methodtwo over \slim is high, the additional time cost is an acceptable trade-off.
Furthermore, although we use the open-source LLM LLama 3.1 in this paper, which eliminate LLM token costs, we also report the average token count required to integrate an integrable set for reference—approximately 476 tokens. Consequently, to complete the task of \taskc on J28, the total inference time and the total token count are 0.56 hours and 2.4 million tokens, respectively. Furthermore, the memory footprint for \methodtwo is 673MB of RAM and 10.4GB of VRAM.

\subsubsection{Trade-off Analysis}
\begin{figure}[tbp]
    \centering
    \begin{subfigure}{0.45\linewidth}
        \centering
        \includegraphics[width=\linewidth]{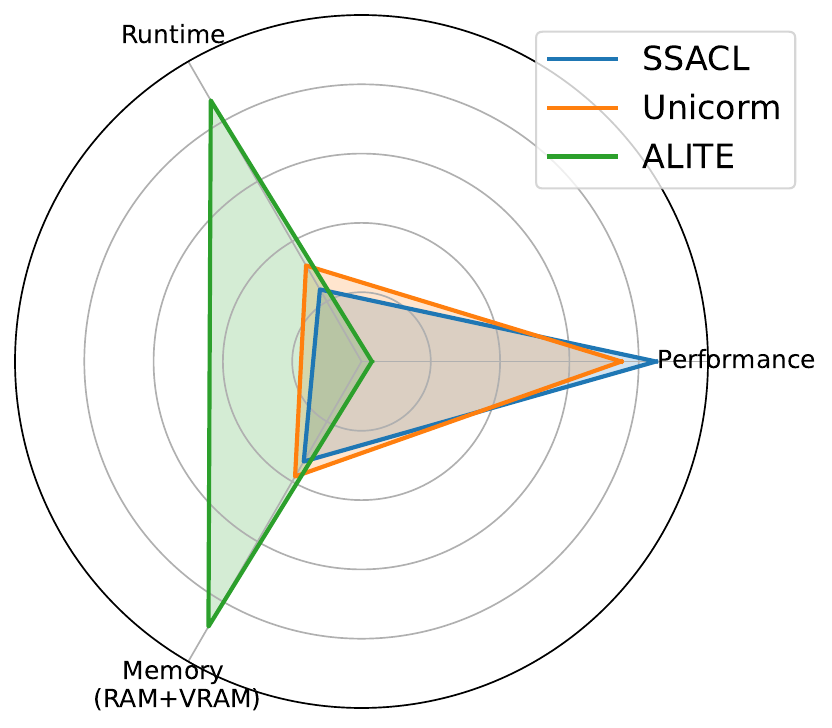}
        \caption{Pairwise integrability judgment}
        \label{Pairwise Integrability Judgment}
    \end{subfigure}
    \hfill
    \begin{subfigure}{0.45\linewidth}
        \centering
        \includegraphics[width=\linewidth]{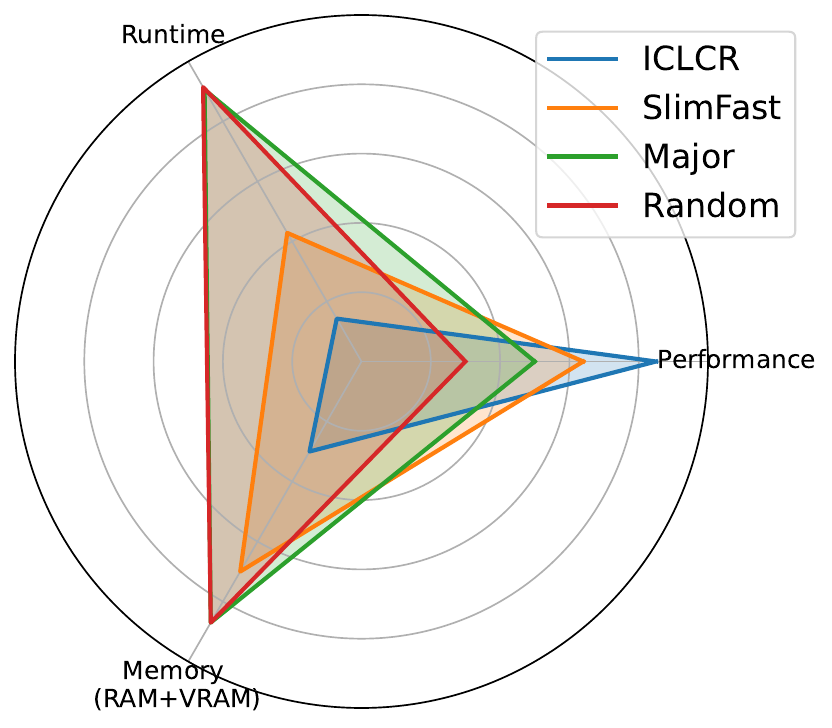}
        \caption{Multi-tuple conflict resolution}
        \label{fig: example2}
    \end{subfigure}
    \caption{Trade-off between performance, run time, and memory footprint among the compared methods.}\label{fig: trade-off}
\end{figure}

We also conduct a trade-off analysis to assess the balance between performance, runtime, and memory footprint across the compared methods. The results, illustrated in Fig.\ref{fig: trade-off} using a spider chart, reveal several key observations. Specifically, in the task of \taska, \methodone demonstrates superior effectiveness compared to Unicorn, albeit with a slight increase in memory usage and runtime. In contrast, while ALITE is highly efficient in terms of speed and memory consumption, its performance is inadequate for practical use. Furthermore, in the \taskb task, \methodtwo~significantly outperforms SlimFast but at the expense of increased memory consumption and computational cost.

\section{Related Work}\label{sec: related work}

\myparagraph{Entity Resolution}
The most relevant work for pairwise integrability judgment is {\em entity resolution}, which determines whether two entities refer to the same entity. 
Existing methods can be divided into four categories: rule-based methods~\cite{vldb_fan2009reasoning,sigmod_singh2017generating,singh2017synthesizing}, crowdsourcing methods~\cite{chai2016cost,chai2018partial,fan2015icrowd,cui2021achieving,yang2018cost}, conventional machine learning (ML)-based methods~\cite{sigkdd_bilenko2003adaptive,sigkdd_tejada2002learning,nips_mccallum2004conditional}, and deep learning (DL)-based methods~\cite{deeper_ebraheem2018distributed,ditto_li2020deep,deepmatcher_mudgal2018deep,dadder_tu2022domain,unicorn_tu2023unicorn,li2023effectiveernew,yao2022hiergat,wang2023sudowoodo,wu2020zeroer,thirumuruganathan2021deepbl}. 

Currently, DL-based methods are considered state-of-the-art and often demonstrate the best overall performance. 
In particular, DeepER~\cite{deeper_ebraheem2018distributed} uses Long Short-Term Memory (LSTM), a type of recurrent neural network, to learn tuple representations using pre-trained word embeddings such as GloVe~\cite{pennington2014glove}.
Then, a multi-layer perceptron (MLP) is used to evaluate the similarity between the two tuple representations.
DeepMatcher~\cite{deepmatcher_mudgal2018deep} categorizes ER problems into three types -- structured ER, textual ER, and ``dirty'' ER -- and provides a comprehensive and systematic solution designs for each.
Ditto~\cite{ditto_li2020deep} leverages domain knowledge and data augmentation, to futher enhance the performance of entity resolution. EMBDI~\cite{cappuzzovldb2020embeddingsheterogeneous} represents the relational database as a compact graph and learns local embeddings that are effective for data integration tasks including entity resolution.
DADER~\cite{dadder_tu2022domain} systematically investigates the problem of domain adaption for entity resolution, addressing scenarios where an ER model trained from a source dataset is applied to a target dataset using little or no labeled data.
Unicorn~\cite{unicorn_tu2023unicorn} introduces a unified framework for entity resolution and five other data matching tasks, and also proposes a multi-task learning framework that enables task performance to be improved holistically.
There are also  several methods that leverage graph neural networks (GNNs) to capture relationships between attributes and entities, thereby improving model performance.
Specifically, HierGAT~\cite{yao2022hiergat} uses contextual embeddings to derive more accurate tuple representations and employs a hierarchical graph attention transformer network to automatically identify the most discriminative words and attributes within a tuple.
On the other hand, FlexER~\cite{genossar2023flexer} tackles universal entity resolution tasks with contemporary solutions to address multiple-intent entity resolution. FlexER formulates the problem as a multi-label classification task. It combines intent-based representations of tuple pairs using a multiplex graph representation, which serves as input to a GNN. By learning intent representations, FlexER enhances performance across multiple resolution problems.

Apart from these supervised ER methods which require high-quality labeled data to train the models, there is also work that investigates how to train an effective machine learning model using limited labeled data in an unsupervised manner.
ZeroER~\cite{wu2020zeroer} tackles Entity Resolution without labeled examples, matching the performance of supervised methods using using Gaussian Mixture Models and adaptive regularization. 
It also incorporates transitivity into its generative model, enhancing accuracy on benchmark ER datasets.
Sudowoodo~\cite{wang2023sudowoodo} is a versatile data integration and preparation framework using contrastive representation learning to perform tasks like entity resolution and error correction without labeled data.
It learns similarity-aware data representations, which can be fine-tuned using a minimal number of labels to achieve state-of-the-art results on various data integration and preparation tasks.

\par While entity resolution is similar to pairwise integrability judgments, there are several important distinctions: 
(1) As introduced in Sec.~\ref{sec: problem formulation}, pairwise integrability judgments represent a broader concept than entity resolution, since two or more tuples that do not refer to the same entity can still be integrable.
(2) Pairwise integrability judgments match tuples containing {\em representative noise}, which include issues such as typographical errors and semantic equivalence. 
(3) Entities in entity resolution are typically derived from information that is not ambiguous, whereas the tuples in our task usually involve missing values -- a common occurrence in large data lakes, especially when using outer union operations.
This disparity demands a different type of comparison at the attribute-level for two tuples, as opposed to a tuple-level assessment.

\myparagraph{Data Fusion} Data fusion~\cite{informationfusion_meng2020survey,csur_bleiholder2009data,vldb_df_azzalini2023enhancing} determines how to merge multiple tuples, which map to the same entity but originate from other resources, into a unified and comprehensive representation.
At the core of data fusion lies the challenge of addressing conflicts when multiple values exist for a particular attribute.
Although a few simple, yet intuitive methods, such as using the mean or median values for numerical values or using majority voting for categorical data can be employed, they often reduce the quality of the data produced. 

\par Existing solutions~\cite{sigmod_df_rekatsinas2017slimfast,broelemann2017ltd,xiao2016towards,vldb_df_li2014confidence} on data fusion predominantly rely on the concept of ``truth discovery'', which involves assessing the reliability of each data source and selecting the most reliable value.
Generally speaking, these methods can be categorized into three groups: probability-based methods~\cite{zhang2016iatd,fang2017truthmulti,lin2018domainawaremtd}, optimization-based methods~\cite{wang2017discovering,li2016conflicts}, and machine learning (ML)-based methods~\cite{xiao2016towards,broelemann2017ltd,sigmod_df_rekatsinas2017slimfast}.
Among these the methods, ML-based truth discovery methods achieve the highest performance.
Specifically, LTD-RBM~\cite{broelemann2017ltd} proposes a novel truth discovery method based on a Restricted Boltzmann Machine (RBM), which supports both continuous and discret values.
LTD-RBM also provides an effective and robust inference procedure based on Contrastive Divergence and Gibbs Sampling.
ETCIBoot~\cite{xiao2016towards} enhances traditional truth discovery techniques by providing confidence intervals with the truth estimates, which can be used to ensure that the judgments are comparable.
ETCIBoot involves updating source weights, estimating truth, and constructing confidence intervals using the bootstrap.
This approach is particularly beneficial in scenarios with varying data density, ensuring more reliable and informative truth discovery.
SLiMFast~\cite{sigmod_df_rekatsinas2017slimfast} frames Data Fusion as a statistical learning problem using discriminative probabilistic models.
It includes components for compilation, optimization, and data fusion, focusing on estimating data source accuracy and predicting true values using statistical learning and probabilistic inference.
Uniquely, SLiMFast combines cross-source conflicts with domain-specific features to improve source accuracy estimation.
\par However, estimating source reliability for truth discovery methods heavily depends on labeled data, metadata, or domain knowledge.
Unfortunately, such data is usually rare in data lakes, which motivates us to develop an effective method using limited labeled data.

\section{Conclusion}
In this paper, we solve three core tasks for data integration in data lake tables, \taska, \taskb, and \taskc.
To solve the task~\taska, we develop a binary classifier, which is able to determine whether any two tuples should be integrated even when they are semantically equivalent, or contain typographical errors.
Then we cast the problem of \taskb to finding maximal cliques or densely connected subgraphs, i.e., communities, in a graph, and explore a collection of representative algorithms to solve it. 
For the task~\taskc, we propose a novel in-context learning (ICL)-based algorithm, which enables us to leverage the extensive knowledge embedded in pretrained large language models to make predictions. 
Finally, our methods show promising performance improvements using limited labeled data. 

Note that this work primarily focuses on ``single-truth'' conflict resolution. 
Due to the one-to-many join relationships among input tables, scenarios may exist where multiple valid resolutions are possible, which will be considered in future work.

\section{Acknowledgment}
Zhifeng Bao is supported in part by ARC DP240101211 and FT240100832.


\begin{thebibliography}{10}

\bibitem{github}
Source code.
\newblock \url{https://github.com/rmitbggroup/Robin}.

\bibitem{abedjan2016detecting}
Ziawasch Abedjan, Xu~Chu, Dong Deng, Raul~Castro Fernandez, Ihab~F Ilyas, Mourad Ouzzani, Paolo Papotti, Michael Stonebraker, and Nan Tang.
\newblock Detecting data errors: Where are we and what needs to be done?
\newblock {\em VLDB 2016}, 9(12):993--1004, 2016.

\bibitem{akyurek2022learningicl}
Ekin Aky{\"u}rek, Dale Schuurmans, Jacob Andreas, Tengyu Ma, and Denny Zhou.
\newblock What learning algorithm is in-context learning? investigations with linear models.
\newblock In {\em ICLR 2022}, 2022.

\bibitem{armandpour2019robustnegativesampling}
Mohammadreza Armandpour, Patrick Ding, Jianhua Huang, and Xia Hu.
\newblock Robust negative sampling for network embedding.
\newblock In {\em AAAI 2019}, volume~33, pages 3191--3198, 2019.

\bibitem{awasthi2022morenegativesampling}
Pranjal Awasthi, Nishanth Dikkala, and Pritish Kamath.
\newblock Do more negative samples necessarily hurt in contrastive learning?
\newblock In {\em ICML 2022}, pages 1101--1116. PMLR, 2022.

\bibitem{vldb_df_azzalini2023enhancing}
Fabio Azzalini, Davide Piantella, Emanuele Rabosio, and Letizia Tanca.
\newblock Enhancing domain-aware multi-truth data fusion using copy-based source authority and value similarity.
\newblock {\em The VLDB Journal}, 32(3):475--500, 2023.

\bibitem{bai2023qwen2}
Jinze Bai, Shuai Bai, Yunfei Chu, Zeyu Cui, Kai Dang, Xiaodong Deng, Yang Fan, Wenbin Ge, Yu~Han, Fei Huang, et~al.
\newblock Qwen technical report.
\newblock {\em arXiv preprint arXiv:2309.16609}, 2023.

\bibitem{csur2022da}
Markus Bayer, Marc-Andr{\'e} Kaufhold, and Christian Reuter.
\newblock A survey on data augmentation for text classification.
\newblock {\em ACM Computing Surveys}, 55(7):1--39, 2022.

\bibitem{sigkdd_bilenko2003adaptive}
Mikhail Bilenko and Raymond~J Mooney.
\newblock Adaptive duplicate detection using learnable string similarity measures.
\newblock In {\em SIGKDD 2003}, pages 39--48, 2003.

\bibitem{csur_bleiholder2009data}
Jens Bleiholder and Felix Naumann.
\newblock Data fusion.
\newblock {\em ACM computing surveys (CSUR)}, 41(1):1--41, 2009.

\bibitem{bleiholder2010subsumption}
Jens Bleiholder, Sascha Szott, Melanie Herschel, Frank Kaufer, and Felix Naumann.
\newblock Subsumption and complementation as data fusion operators.
\newblock In {\em EDBT 2010}, pages 513--524, 2010.

\bibitem{louvain}
Vincent~D Blondel, Jean-Loup Guillaume, Renaud Lambiotte, and Etienne Lefebvre.
\newblock Fast unfolding of communities in large networks.
\newblock {\em Journal of statistical mechanics: theory and experiment}, 2008(10):P10008, 2008.

\bibitem{bojanowski2017fasttext}
Piotr Bojanowski, Edouard Grave, Armand Joulin, and Tomas Mikolov.
\newblock Enriching word vectors with subword information.
\newblock {\em TACL}, 5:135--146, 2017.

\bibitem{lagrange-multiplier-book}
Stephen Boyd and Lieven Vandenberghe.
\newblock {\em Convex Optimization}.
\newblock Cambridge University Press, Cambridge, UK, 2004.

\bibitem{brinkmann2023iclexample}
Alexander Brinkmann, Roee Shraga, and Christian Bizer.
\newblock Product attribute value extraction using large language models.
\newblock {\em arXiv preprint arXiv:2310.12537}, 2023.

\bibitem{brinkmann2024CL1}
Alexander Brinkmann, Roee Shraga, and Christina Bizer.
\newblock Sc-block: Supervised contrastive blocking within entity resolution pipelines.
\newblock In {\em ESWC}, pages 121--142. Springer, 2024.

\bibitem{broelemann2017ltd}
Klaus Broelemann, Thomas Gottron, and Gjergji Kasneci.
\newblock Ltd-rbm: Robust and fast latent truth discovery using restricted boltzmann machines.
\newblock In {\em ICDE 2017}, pages 143--146. IEEE, 2017.

\bibitem{bruna2017community}
Joan Bruna and X~Li.
\newblock Community detection with graph neural networks.
\newblock {\em stat}, 1050:27, 2017.

\bibitem{cappuzzo2024retrieve}
Riccardo Cappuzzo, Aimee Coelho, Felix Lefebvre, Paolo Papotti, and Gael Varoquaux.
\newblock Retrieve, merge, predict: Augmenting tables with data lakes.
\newblock {\em arXiv preprint arXiv:2402.06282}, 2024.

\bibitem{cappuzzovldb2020embeddingsheterogeneous}
Riccardo Cappuzzo, Paolo Papotti, and Saravanan Thirumuruganathan.
\newblock Creating embeddings of heterogeneous relational datasets for data integration tasks.
\newblock In {\em SIGMOD 2020}, pages 1335--1349, 2020.

\bibitem{chai2016cost}
Chengliang Chai, Guoliang Li, Jian Li, Dong Deng, and Jianhua Feng.
\newblock Cost-effective crowdsourced entity resolution: A partial-order approach.
\newblock In {\em SIGMOD 2016}, pages 969--984, 2016.

\bibitem{chai2018partial}
Chengliang Chai, Guoliang Li, Jian Li, Dong Deng, and Jianhua Feng.
\newblock A partial-order-based framework for cost-effective crowdsourced entity resolution.
\newblock {\em The VLDB Journal}, 27:745--770, 2018.

\bibitem{cui2021achieving}
Lizhen Cui, Jing Chen, Wei He, Hui Li, Wei Guo, and Zhiyuan Su.
\newblock Achieving approximate global optimization of truth inference for crowdsourcing microtasks.
\newblock {\em Data Science and Engineering}, 6(3):294--309, 2021.

\bibitem{devlin2019bert}
Jacob Devlin, Ming-Wei Chang, Kenton Lee, and Kristina Toutanova.
\newblock Bert: Pre-training of deep bidirectional transformers for language understanding.
\newblock In {\em NAACL 2019}, pages 4171--4186, 2019.

\bibitem{vldb_df_dong2009data}
Xin~Luna Dong and Felix Naumann.
\newblock Data fusion: resolving data conflicts for integration.
\newblock {\em VLDB 2009}, 2(2):1654--1655, 2009.

\bibitem{dubey2024llama3}
Abhimanyu Dubey, Abhinav Jauhri, Abhinav Pandey, Abhishek Kadian, Ahmad Al-Dahle, Aiesha Letman, Akhil Mathur, Alan Schelten, Amy Yang, Angela Fan, et~al.
\newblock The llama 3 herd of models.
\newblock {\em arXiv preprint arXiv:2407.21783}, 2024.

\bibitem{deeper_ebraheem2018distributed}
Muhammad Ebraheem, Saravanan Thirumuruganathan, Shafiq Joty, Mourad Ouzzani, and Nan Tang.
\newblock Distributed representations of tuples for entity resolution.
\newblock {\em VLDB 2018}, 11(11):1454--1467, 2018.

\bibitem{edunov2018understanding}
Sergey Edunov, Myle Ott, Michael Auli, and David Grangier.
\newblock Understanding back-translation at scale.
\newblock In {\em EMNLP 2018}, pages 489--500, 2018.

\bibitem{fan2023tabletsnew}
Grace Fan, Jin Wang, Yuliang Li, and Ren{\'e}e~J Miller.
\newblock Table discovery in data lakes: State-of-the-art and future directions.
\newblock In {\em Companion of SIGMOD 2023}, pages 69--75, 2023.

\bibitem{starmie_fan2022semantics}
Grace Fan, Jin Wang, Yuliang Li, Dan Zhang, and Ren{\'e}e~J Miller.
\newblock Semantics-aware dataset discovery from data lakes with contextualized column-based representation learning.
\newblock {\em VLDB 2023}, 16(7):1726--1739, 2023.

\bibitem{fan2015icrowd}
Ju~Fan, Guoliang Li, Beng~Chin Ooi, Kian-lee Tan, and Jianhua Feng.
\newblock icrowd: An adaptive crowdsourcing framework.
\newblock In {\em SIGMOD 2015}, pages 1015--1030, 2015.

\bibitem{vldb_fan2009reasoning}
Wenfei Fan, Xibei Jia, Jianzhong Li, and Shuai Ma.
\newblock Reasoning about record matching rules.
\newblock {\em VLDB 2009}, 2(1):407--418, 2009.

\bibitem{fang2017truthmulti}
Xiu~Susie Fang.
\newblock Truth discovery from conflicting multi-valued objects.
\newblock In {\em WWW 2017 Companion}, pages 711--715, 2017.

\bibitem{feng2021dafornlp}
Steven~Y Feng, Varun Gangal, Jason Wei, Sarath Chandar, Soroush Vosoughi, Teruko Mitamura, and Eduard Hovy.
\newblock A survey of data augmentation approaches for nlp.
\newblock In {\em Findings of ACL 2021}, pages 968--988, 2021.

\bibitem{genossar2023flexer}
Bar Genossar, Roee Shraga, and Avigdor Gal.
\newblock Flexer: flexible entity resolution for multiple intents.
\newblock {\em SIGMOD 2023}, 1(1):1--27, 2023.

\bibitem{goodfellowiclr2014explaining}
Ian~J Goodfellow, Jonathon Shlens, and Christian Szegedy.
\newblock Explaining and harnessing adversarial examples.
\newblock {\em arXiv preprint arXiv:1412.6572}, 2014.

\bibitem{constance_hai2016constance}
Rihan Hai, Sandra Geisler, and Christoph Quix.
\newblock Constance: An intelligent data lake system.
\newblock In {\em SIGMOD 2016}, pages 2097--2100, 2016.

\bibitem{he2020deberta}
Pengcheng He, Xiaodong Liu, Jianfeng Gao, and Weizhu Chen.
\newblock Deberta: Decoding-enhanced bert with disentangled attention.
\newblock {\em arXiv preprint arXiv:2006.03654}, 2020.

\bibitem{ilyasnlps2019adversarial}
Andrew Ilyas, Shibani Santurkar, Dimitris Tsipras, Logan Engstrom, Brandon Tran, and Aleksander Madry.
\newblock Adversarial examples are not bugs, they are features.
\newblock {\em NIPS 2019}, 32, 2019.

\bibitem{jiang2023mistral}
Albert~Q Jiang, Alexandre Sablayrolles, Arthur Mensch, Chris Bamford, Devendra~Singh Chaplot, Diego de~las Casas, Florian Bressand, Gianna Lengyel, Guillaume Lample, Lucile Saulnier, et~al.
\newblock Mistral 7b.
\newblock {\em arXiv preprint arXiv:2310.06825}, 2023.

\bibitem{kenton2019bert}
Jacob Devlin Ming-Wei~Chang Kenton and Lee~Kristina Toutanova.
\newblock Bert: Pre-training of deep bidirectional transformers for language understanding.
\newblock In {\em NAACL 2019}, pages 4171--4186, 2019.

\bibitem{santos_khatiwada2023santos}
Aamod Khatiwada, Grace Fan, Roee Shraga, Zixuan Chen, Wolfgang Gatterbauer, Ren{\'e}e~J Miller, and Mirek Riedewald.
\newblock Santos: Relationship-based semantic table union search.
\newblock {\em SIGMOD 2023}, 1(1):1--25, 2023.

\bibitem{alite_khatiwada2022integrating}
Aamod Khatiwada, Roee Shraga, Wolfgang Gatterbauer, and Ren{\'e}e~J Miller.
\newblock Integrating data lake tables.
\newblock {\em VLDB 2022}, 16(4):932--945, 2022.

\bibitem{khatiwada2025fuzzy}
Aamod Khatiwada, Roee Shraga, and Ren{\'e}e~J Miller.
\newblock Fuzzy integration of data lake tables.
\newblock {\em arXiv preprint arXiv:2501.09211}, 2025.

\bibitem{nips_scl_khosla2020supervised}
Prannay Khosla, Piotr Teterwak, Chen Wang, Aaron Sarna, Yonglong Tian, Phillip Isola, Aaron Maschinot, Ce~Liu, and Dilip Krishnan.
\newblock Supervised contrastive learning.
\newblock {\em NIPS 2020}, 33:18661--18673, 2020.

\bibitem{kingma2014adam}
DP~Kingma.
\newblock Adam: a method for stochastic optimization.
\newblock In {\em ICLR 2014}, 2014.

\bibitem{typo_github}
Ranvijay Kumar.
\newblock Typo - a github repository, 2025.
\newblock Accessed: 2025-01-18.

\bibitem{vldb_df_li2014confidence}
Qi~Li, Yaliang Li, Jing Gao, Lu~Su, Bo~Zhao, Murat Demirbas, Wei Fan, and Jiawei Han.
\newblock A confidence-aware approach for truth discovery on long-tail data.
\newblock {\em VLDB 2015}, 8(4):425--436, 2014.

\bibitem{voting_li2012truth}
Xian Li, Xin~Luna Dong, Kenneth Lyons, Weiyi Meng, and Divesh Srivastava.
\newblock Truth finding on the deep web: Is the problem solved?
\newblock {\em VLDB 2013}, 6(2), 2012.

\bibitem{li2016conflicts}
Yaliang Li, Qi~Li, Jing Gao, Lu~Su, Bo~Zhao, Wei Fan, and Jiawei Han.
\newblock Conflicts to harmony: A framework for resolving conflicts in heterogeneous data by truth discovery.
\newblock {\em TKDE}, 28(8):1986--1999, 2016.

\bibitem{li2023effectiveernew}
Yuliang Li, Jinfeng Li, Yoshi Suhara, AnHai Doan, and Wang-Chiew Tan.
\newblock Effective entity matching with transformers.
\newblock {\em The VLDB Journal}, pages 1--21, 2023.

\bibitem{ditto_li2020deep}
Yuliang Li, Jinfeng Li, Yoshihiko Suhara, AnHai Doan, and Wang-Chiew Tan.
\newblock Deep entity matching with pre-trained language models.
\newblock {\em VLDB 2021}, 14(1):50--60, 2020.

\bibitem{lin2018domainawaremtd}
Xueling Lin and Lei Chen.
\newblock Domain-aware multi-truth discovery from conflicting sources.
\newblock {\em VLDB 2018}, 11(5):635--647, 2018.

\bibitem{liu2019roberta}
Yinhan Liu, Myle Ott, Naman Goyal, Jingfei Du, Mandar Joshi, Danqi Chen, Omer Levy, Mike Lewis, Luke Zettlemoyer, and Veselin Stoyanov.
\newblock Roberta: A robustly optimized bert pretraining approach.
\newblock {\em arXiv preprint arXiv:1907.11692}, 2019.

\bibitem{nips_mccallum2004conditional}
Andrew McCallum and Ben Wellner.
\newblock Conditional models of identity uncertainty with application to noun coreference.
\newblock {\em NIPS 2004}, 17, 2004.

\bibitem{informationfusion_meng2020survey}
Tong Meng, Xuyang Jing, Zheng Yan, and Witold Pedrycz.
\newblock A survey on machine learning for data fusion.
\newblock {\em Information Fusion}, 57:115--129, 2020.

\bibitem{mikolov2013word2vec}
Tomas Mikolov, Ilya Sutskever, Kai Chen, Greg~S Corrado, and Jeff Dean.
\newblock Distributed representations of words and phrases and their compositionality.
\newblock {\em NIPS 2013}, 26, 2013.

\bibitem{miller1995wordnet}
George~A Miller.
\newblock Wordnet: a lexical database for english.
\newblock {\em Communications of the ACM}, 38(11):39--41, 1995.

\bibitem{min2022rethinkingicl}
Sewon Min, Xinxi Lyu, Ari Holtzman, Mikel Artetxe, Mike Lewis, Hannaneh Hajishirzi, and Luke Zettlemoyer.
\newblock Rethinking the role of demonstrations: What makes in-context learning work?
\newblock In {\em EMNLP 2022}, pages 11048--11064, 2022.

\bibitem{deepmatcher_mudgal2018deep}
Sidharth Mudgal, Han Li, Theodoros Rekatsinas, AnHai Doan, Youngchoon Park, Ganesh Krishnan, Rohit Deep, Esteban Arcaute, and Vijay Raghavendra.
\newblock Deep learning for entity matching: A design space exploration.
\newblock In {\em SIGMOD 2018}, pages 19--34, 2018.

\bibitem{vldb_data_lake_management_nargesian2019data}
Fatemeh Nargesian, Erkang Zhu, Ren{\'e}e~J Miller, Ken~Q Pu, and Patricia~C Arocena.
\newblock Data lake management: challenges and opportunities.
\newblock {\em VLDB 2019}, 12(12):1986--1989, 2019.

\bibitem{nargesian2018tabletsnew}
Fatemeh Nargesian, Erkang Zhu, Ken~Q Pu, and Ren{\'e}e~J Miller.
\newblock Table union search on open data.
\newblock {\em VLDB 2018}, 11(7):813--825, 2018.

\bibitem{newman2004NGalg}
M.~E.~J. Newman and M.~Girvan.
\newblock Finding and evaluating community structure in networks.
\newblock {\em Physical Review E}, 69(2):026113, 2004.

\bibitem{newman2013spectral}
Mark~EJ Newman.
\newblock Spectral methods for community detection and graph partitioning.
\newblock {\em Physical Review E}, 88(4):042822, 2013.

\bibitem{ng2001spectral}
Andrew~Y. Ng, Michael~I. Jordan, and Yair Weiss.
\newblock On spectral clustering: Analysis and an algorithm.
\newblock In {\em NIPS 2001}, pages 849--856, 2001.

\bibitem{NoceWrig06}
Jorge Nocedal and Stephen~J. Wright.
\newblock {\em Numerical Optimization}.
\newblock Springer, New York, NY, USA, 2e edition, 2006.

\bibitem{peng2022garf}
Jinfeng Peng, Derong Shen, Nan Tang, Tieying Liu, Yue Kou, Tiezheng Nie, Hang Cui, and Ge~Yu.
\newblock Self-supervised and interpretable data cleaning with sequence generative adversarial networks.
\newblock {\em VLDB 2023}, 16(3):433--446, 2022.

\bibitem{pennington2014glove}
Jeffrey Pennington, Richard Socher, and Christopher~D Manning.
\newblock Glove: Global vectors for word representation.
\newblock In {\em EMNLP 2014}, pages 1532--1543, 2014.

\bibitem{regneri2007bronalg}
Michaela Regneri.
\newblock Finding all cliques of an undirected graph.
\newblock In {\em Seminar current trends in IE WS jun}, 2007.

\bibitem{sigmod_df_rekatsinas2017slimfast}
Theodoros Rekatsinas, Manas Joglekar, Hector Garcia-Molina, Aditya Parameswaran, and Christopher R{\'e}.
\newblock Slimfast: Guaranteed results for data fusion and source reliability.
\newblock In {\em SIGMOD 2017}, pages 1399--1414, 2017.

\bibitem{rosvall2008infomap}
Martin Rosvall and Carl~T Bergstrom.
\newblock Maps of random walks on complex networks reveal community structure.
\newblock {\em PNAS}, 105(4):1118--1123, 2008.

\bibitem{sanh2019distilbert}
Victor Sanh, Lysandre Debut, Julien Chaumond, and Thomas Wolf.
\newblock Distilbert, a distilled version of bert: smaller, faster, cheaper and lighter.
\newblock {\em arXiv preprint arXiv:1910.01108}, 2019.

\bibitem{shorten2021text}
Connor Shorten, Taghi~M Khoshgoftaar, and Borko Furht.
\newblock Text data augmentation for deep learning.
\newblock {\em Journal of big Data}, 8:1--34, 2021.

\bibitem{simonini2018schemaagnostic1}
Giovanni Simonini, George Papadakis, Themis Palpanas, and Sonia Bergamaschi.
\newblock Schema-agnostic progressive entity resolution.
\newblock {\em TKDE}, 31(6):1208--1221, 2018.

\bibitem{sigmod_singh2017generating}
Rohit Singh, Vamsi Meduri, Ahmed Elmagarmid, Samuel Madden, Paolo Papotti, Jorge-Arnulfo Quian{\'e}-Ruiz, Armando Solar-Lezama, and Nan Tang.
\newblock Generating concise entity matching rules.
\newblock In {\em SIGMOD 2017}, pages 1635--1638, 2017.

\bibitem{singh2017synthesizing}
Rohit Singh, Venkata~Vamsikrishna Meduri, Ahmed Elmagarmid, Samuel Madden, Paolo Papotti, Jorge-Arnulfo Quian{\'e}-Ruiz, Armando Solar-Lezama, and Nan Tang.
\newblock Synthesizing entity matching rules by examples.
\newblock {\em VLDB 2018}, 11(2):189--202, 2017.

\bibitem{sigkdd_tejada2002learning}
Sheila Tejada, Craig~A Knoblock, and Steven Minton.
\newblock Learning domain-independent string transformation weights for high accuracy object identification.
\newblock In {\em SIGKDD 2002}, pages 350--359, 2002.

\bibitem{teong2020schemaagnostic2}
Kai-Sheng Teong, Lay-Ki Soon, and Tin~Tin Su.
\newblock Schema-agnostic entity matching using pre-trained language models.
\newblock In {\em CIKM 2020}, pages 2241--2244, 2020.

\bibitem{thirumuruganathan2021deepbl}
Saravanan Thirumuruganathan, Han Li, Nan Tang, Mourad Ouzzani, Yash Govind, Derek Paulsen, Glenn Fung, and AnHai Doan.
\newblock Deep learning for blocking in entity matching: a design space exploration.
\newblock {\em VLDB 2021}, 14(11):2459--2472, 2021.

\bibitem{touvron2023llama}
Hugo Touvron, Thibaut Lavril, Gautier Izacard, Xavier Martinet, Marie-Anne Lachaux, Timoth{\'e}e Lacroix, Baptiste Rozi{\`e}re, Naman Goyal, Eric Hambro, Faisal Azhar, et~al.
\newblock Llama: Open and efficient foundation language models.
\newblock {\em arXiv preprint arXiv:2302.13971}, 2023.

\bibitem{dadder_tu2022domain}
Jianhong Tu, Ju~Fan, Nan Tang, Peng Wang, Chengliang Chai, Guoliang Li, Ruixue Fan, and Xiaoyong Du.
\newblock Domain adaptation for deep entity resolution.
\newblock In {\em SIGMOD 2022}, pages 443--457, 2022.

\bibitem{unicorn_tu2023unicorn}
Jianhong Tu, Ju~Fan, Nan Tang, Peng Wang, Guoliang Li, Xiaoyong Du, Xiaofeng Jia, and Song Gao.
\newblock Unicorn: A unified multi-tasking model for supporting matching tasks in data integration.
\newblock {\em SIGMOD 2023}, 1(1):1--26, 2023.

\bibitem{vaswani2017attention}
Ashish Vaswani, Noam Shazeer, Niki Parmar, Jakob Uszkoreit, Llion Jones, Aidan~N Gomez, {\L}ukasz Kaiser, and Illia Polosukhin.
\newblock Attention is all you need.
\newblock {\em NIPS 2017}, 30, 2017.

\bibitem{wang2023sudowoodo}
Runhui Wang, Yuliang Li, and Jin Wang.
\newblock Sudowoodo: Contrastive self-supervised learning for multi-purpose data integration and preparation.
\newblock In {\em ICDE 2023}, pages 1502--1515. IEEE, 2023.

\bibitem{wang2017discovering}
Yaqing Wang, Fenglong Ma, Lu~Su, and Jing Gao.
\newblock Discovering truths from distributed data.
\newblock In {\em ICDM 2017}, pages 505--514. IEEE, 2017.

\bibitem{wu2020zeroer}
Renzhi Wu, Sanya Chaba, Saurabh Sawlani, Xu~Chu, and Saravanan Thirumuruganathan.
\newblock Zeroer: Entity resolution using zero labeled examples.
\newblock In {\em SIGMOD 2020}, pages 1149--1164, 2020.

\bibitem{xiao2016towards}
Houping Xiao, Jing Gao, Qi~Li, Fenglong Ma, Lu~Su, Yunlong Feng, and Aidong Zhang.
\newblock Towards confidence in the truth: A bootstrapping based truth discovery approach.
\newblock In {\em SIGKDD 2016}, pages 1935--1944, 2016.

\bibitem{xie2021explanationicl}
Sang~Michael Xie, Aditi Raghunathan, Percy Liang, and Tengyu Ma.
\newblock An explanation of in-context learning as implicit bayesian inference.
\newblock In {\em ICLR 2021}, 2021.

\bibitem{yang2018cost}
Jingru Yang, Ju~Fan, Zhewei Wei, Guoliang Li, Tongyu Liu, and Xiaoyong Du.
\newblock Cost-effective data annotation using game-based crowdsourcing.
\newblock {\em VLDB 2018}, 12(1):57--70, 2018.

\bibitem{yang2019xlnet}
Zhilin Yang, Zihang Dai, Yiming Yang, Jaime Carbonell, Ruslan Salakhutdinov, and Quoc~V. Le.
\newblock Xlnet: Generalized autoregressive pretraining for language understanding.
\newblock In {\em NIPS 2019}, pages 5753--5763, 2019.

\bibitem{yao2022hiergat}
Dezhong Yao, Yuhong Gu, Gao Cong, Hai Jin, and Xinqiao Lv.
\newblock Entity resolution with hierarchical graph attention networks.
\newblock In {\em SIGMOD 2022}, pages 429--442, 2022.

\bibitem{zhang2016iatd}
Hengtong Zhang, Qi~Li, Fenglong Ma, Houping Xiao, Yaliang Li, Jing Gao, and Lu~Su.
\newblock Influence-aware truth discovery.
\newblock In {\em CIKM 2016}, pages 851--860, 2016.

\bibitem{zhangtnnls2019adversarial}
Jiliang Zhang and Chen Li.
\newblock Adversarial examples: Opportunities and challenges.
\newblock {\em TNNLS}, 31(7):2578--2593, 2019.

\bibitem{sigkdd_df_zhi2015modeling}
Shi Zhi, Bo~Zhao, Wenzhu Tong, Jing Gao, Dian Yu, Heng Ji, and Jiawei Han.
\newblock Modeling truth existence in truth discovery.
\newblock In {\em SIGMOD 2015}, pages 1543--1552, 2015.

\bibitem{zhu2019josietsnew}
Erkang Zhu, Dong Deng, Fatemeh Nargesian, and Ren{\'e}e~J Miller.
\newblock Josie: Overlap set similarity search for finding joinable tables in data lakes.
\newblock In {\em SIGMOD 2019}, pages 847--864, 2019.

\end{thebibliography}

\end{document}